\documentclass[conference]{IEEEtran}
\IEEEoverridecommandlockouts

\usepackage{cite}
\usepackage{amsmath,amssymb,amsfonts}
\usepackage{algorithm}
\usepackage{algorithmicx}
\usepackage{algpseudocode}
\usepackage{graphicx}
\usepackage{subcaption}
\usepackage{textcomp}
\usepackage{url}
\usepackage{xcolor}
\def\BibTeX{{\rm B\kern-.05em{\sc i\kern-.025em b}\kern-.08em
    T\kern-.1667em\lower.7ex\hbox{E}\kern-.125emX}}
\begin{document}
\pagestyle{plain}

\title{Accelerating Locality-Driven Integration in Quantum Chemistry with Block-Structured Matrix Multiplication
\thanks{\IEEEauthorrefmark{1} Equal contribution. Correspondence to: jufusong@zga.edu.cn.}
}


\author{
\IEEEauthorblockN{
Xinran Wei\IEEEauthorrefmark{1}\IEEEauthorrefmark{2}\IEEEauthorrefmark{5},
Yan Pan\IEEEauthorrefmark{1}, 
Fusong Ju\IEEEauthorrefmark{1}\IEEEauthorrefmark{2},
Zehao Zhou\IEEEauthorrefmark{2},
Yihong Zhang\IEEEauthorrefmark{6}\IEEEauthorrefmark{2},
Lin Huang\IEEEauthorrefmark{4}, \\
Jianwei Zhu\IEEEauthorrefmark{2},
Jia Zhang\IEEEauthorrefmark{4},
Huanhuan Xia\IEEEauthorrefmark{2}\IEEEauthorrefmark{5},
Bin Shao\IEEEauthorrefmark{2}\IEEEauthorrefmark{5},
Tao Qin\IEEEauthorrefmark{2}
}
\IEEEauthorblockA{\IEEEauthorrefmark{2}\textit{Zhongguancun Academy}, Beijing, China}
\IEEEauthorblockA{\IEEEauthorrefmark{5}\textit{Zhongguancun Institute of Artificial Intelligence}, Beijing, China}
\IEEEauthorblockA{\IEEEauthorrefmark{4}\textit{IQuest Research}, Beijing, China}
\IEEEauthorblockA{\IEEEauthorrefmark{6}\textit{University of Science and Technology of China}, Hefei, Anhui, China}
}

\maketitle

\begin{abstract}
Locality-driven integration is a pervasive computational pattern in quantum chemistry, arising whenever spatially localized basis functions interact through numerical quadrature or integral screening. The dominant matrix multiplications in these tasks exhibit dynamic, structured sparsity driven by spatial locality, posing significant challenges for both dense batched kernels and generic sparse formats on GPUs. We present KerneLDI, a GPU-oriented framework that addresses this regime by co-designing data layout, screening logic, and matrix-computation operators to realize block-structured matrix multiplication for locality-driven integration. KerneLDI reorganizes operand matrices into a unified block-filtered representation that retains only spatially relevant blocks, and executes the resulting contractions with customized dense block multipliers that adapt proven dense-matmul optimizations to retained block pairs. We develop and evaluate KerneLDI on exchange--correlation (EXC) integration in Kohn--Sham density functional theory, a representative and computationally critical instance of this pattern. Across diverse molecular systems, KerneLDI preserves numerical accuracy while delivering up to 10$\times$ speedup for EXC evaluation over a dense GPU baseline, scales favorably with increasing system size and multi-GPU parallelism, accelerates end-to-end self-consistent field calculations, and yields nearly 6$\times$ throughput improvement for ab initio molecular dynamics.

\end{abstract}

\begin{IEEEkeywords}
GPU computing, density functional theory, locality-driven integration, structured sparsity, block-structured matrix multiplication.
\end{IEEEkeywords}

\section{Introduction}

Matrix operations are a foundational abstraction for modern high-performance computing. Dense matrix multiplication and tensor contraction kernels sit at the core of numerical simulation, machine learning, and scientific computing because they expose regular data reuse and map well to accelerator hardware~\cite{nickolls2010gpu,volkov2008benchmarking,cublas}. Over the past decade, GPU software stacks and hardware-aware libraries have made dense matrix workloads increasingly efficient through hierarchical tiling, shared-memory staging, register-level accumulation, batching, kernel fusion, and hardware-specialized memory pipelines~\cite{cuda2019programming}. In parallel, the HPC community has also made substantial progress on sparse and irregular matrix computation through compressed sparse formats, block-sparse representations, graph and hypergraph reorderings, task-based scheduling, and specialized sparse kernels such as SpMV and SpMM~\cite{bell2009implementing,cusparse,filippone2017sparse,bulucc2012parallel,yang2018design}. These advances have greatly improved the tractability of many sparse linear-algebra workloads, but they do not eliminate the gap between dense-kernel efficiency and irregular scientific matrices.

That gap persists because generic sparse methods are typically most effective when sparsity is sufficiently high, the indexing overhead can be amortized, and the access pattern is stable enough to benefit from reusable data layouts~\cite{filippone2017sparse,davis2011university}. Conversely, forcing irregular data back into dense tiles or batches can recover some locality, but usually at the cost of padding, wasted memory traffic, and strong dependence on preprocessing heuristics~\cite{vuduc2005fast,eberhardt2016optimization}. A difficult middle ground remains: matrices that are neither dense nor extremely sparse, whose structure is irregular but not arbitrary, and whose performance depends on preserving locality without paying the full cost of generic sparse indirection.

Locality-driven integration is a typical workload that falls into this difficult middle ground in quantum chemistry. Exchange--correlation integration in density functional theory, Coulomb density fitting via the resolution-of-identity (RI) approximation~\cite{vahtras1993ri,kussmann2021highly}, Schwarz-screened exact-exchange construction~\cite{laqua2020snlink}, and related workloads all exhibit the shared pattern of locality-driven integration: spatially localized basis functions or grid partitions interact through numerical quadrature or integral screening. Although these tasks are often introduced as grid-based integration procedures, their dominant cost can be reorganized as matrix multiplications of locality-governed retained contributions, typically involving basis-function evaluation, density-dependent weights, and Fock-like accumulation. The resulting workloads are therefore neither dense nor extremely sparse: contributions are retained dynamically according to locality, geometry, screening thresholds, and grid organization, placing them squarely in the middle ground described above.

\begin{figure}[t!]
    \centering
    \includegraphics[width=0.48\textwidth]{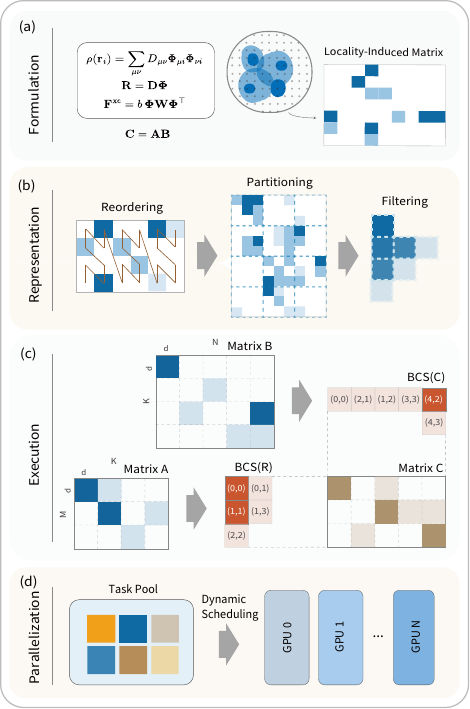}
    \caption{Overview of the KerneLDI workflow. EXC operands generated from basis functions and integration grids are reordered, partitioned, and filtered to produce a compact block-structured representation. Retained blocks are organized in row- and column-oriented BCS layouts and evaluated in a block-pair fashion. Dynamic scheduling distributes tasks across multiple GPUs for parallel execution.}
    \label{fig:2}
\end{figure}

Current GPU implementations of this class of locality-driven integration workloads often regularize their irregular retained patterns into dense batched GEMM kernels. That regularization improves nominal occupancy, but it also introduces padding and wasted memory traffic when heterogeneous basis support is forced into a common batch shape~\cite{manathunga2020parallel,williams2021achieving,wu2025enhancing,stocks2025efficient}. Generic fine-grained sparse formats are also a poor fit, because their metadata overhead and irregular access patterns make it difficult to preserve locality and arithmetic intensity on GPUs~\cite{bell2009implementing,filippone2017sparse}. Together, these limitations indicate that locality-driven integration requires customized data structures and matrix-computation operators, rather than a fallback to either dense batching or generic sparse execution.

This observation points to a co-design problem across data layout, screening logic, and computational operators. Rather than treating sparsity only as a preprocessing step before reverting to dense kernels, the representation and execution model should be designed together to preserve the arithmetic structure of locality-driven integration while matching the memory/computation hierarchy of modern accelerators. We therefore present KerneLDI, a GPU-oriented framework for block-structured matrix multiplication in locality-driven integration. It reorganizes operand matrices into a unified block-filtered format and executes the corresponding contractions with customized dense block multipliers. The representation preserves spatial locality while allowing different grid batches to retain different basis blocks without being forced into a common dense shape. The execution model screens block pairs dynamically, improves memory coalescing and arithmetic intensity, and naturally adapts proven dense-matmul optimizations to KerneLDI.

Among typical locality-driven integration tasks, exchange--correlation (EXC) integration in Kohn--Sham density functional theory is both a representative and a scientifically critical case~\cite{becke1988multicenter,treutler1995grid,gill1993grid}. It lies on the critical path of self-consistent field iterations, analytic gradients, and ab initio molecular dynamics, and its cost typically scales as $O(n^3)$ because both the basis size and grid size grow approximately linearly with system size~\cite{williams2021achieving,manathunga2020parallel}. As a result, EXC directly limits the reachable system sizes and simulated timescales of DFT workloads on modern accelerators~\cite{marx2009ab,kresse1996efficient,hutter2014cp2k}. We therefore take EXC as the workload for developing and evaluating our approach in this paper, although the block-filtered representation and execution model are designed to be applicable to locality-driven integration tasks more broadly.

In summary, the main contributions of this work are as follows:
\begin{itemize}
    \item Taking EXC integration as a representative case, we formulate locality-driven integration as a sequence of dynamically filtered matrix multiplications, making the underlying sparsity explicit and amenable to block-structured GPU execution.
    \item We introduce a block-filtered transformation that unifies the major matrix operations of locality-driven integration under a common block-structured representation, enabling matrix multiplication over retained blocks with efficient indexing.
    \item We design GPU dense block multipliers that exploit the block-filtered representation for high-locality, high-intensity execution on retained block pairs, leveraging Tensor Core acceleration and other proven dense-matmul optimizations on the filtered workload.
    \item We develop a dynamic multi-GPU execution model that decomposes filtered EXC workloads into heterogeneous grid-batch tasks and schedules them on demand to mitigate load imbalance across devices.
    \item We demonstrate on EXC that KerneLDI preserves numerical accuracy while delivering up to 10$\times$ speedup for EXC evaluation over a dense GPU baseline, scales favorably with increasing system size and multi-GPU parallelism, accelerates end-to-end self-consistent field calculations, and yields nearly 6$\times$ throughput improvement for ab initio molecular dynamics.
\end{itemize}



\section{Workload Formulation and Challenges}

\subsection{EXC as Matrix Products}

The exchange--correlation (EXC) contribution is a dominant component of Kohn--Sham density functional theory (KS-DFT) workloads because it lies on the critical path of self-consistent field (SCF) iterations, analytic gradients, and \textit{ab initio} molecular dynamics. In practical Gaussian-basis KS-DFT implementations, its dominant cost arises from numerical quadrature over atom-centered integration grids and can be reorganized into matrix products whose structure is determined by basis locality and grid support.

Let $\{(\mathbf r_i, w_i)\}_{i=1}^{N_{\mathrm{grid}}}$ denote the integration grid points and quadrature weights. The EXC contribution to the Fock matrix is given by
\begin{equation}
\begin{aligned}
\mathbf{F}^{\mathrm{xc}}_{\mu\nu}
&=
\int_{\mathbb R^3}
\chi_\mu(\mathbf r)\,
\frac{\partial E_{\mathrm{xc}}[\rho(\mathbf r)]}{\partial \rho(\mathbf r)}\,
\chi_\nu(\mathbf r)\, d\mathbf r \\
&\approx
b \sum_{i=1}^{N_{\mathrm{grid}}}
w_i\, v_i^{\mathrm{xc}}\,
\chi_\mu(\mathbf r_i)\chi_\nu(\mathbf r_i),
\end{aligned}
\label{eq:exc_quadrature}
\end{equation}
where $\chi_\mu(\mathbf r)$ and $\chi_\nu(\mathbf r)$ are basis functions, $v_i^{\mathrm{xc}}$ denotes the exchange--correlation potential evaluated on grid point $\mathbf r_i$, and $b$ is a constant prefactor determined by the spin treatment ($b=1$ for restricted, $b=2$ for unrestricted calculations; see the Appendix). Although Eq.~\eqref{eq:exc_quadrature} is written as a weighted summation over grid points, its dominant cost can be recast into a small number of matrix operations.

We therefore introduce the basis-on-grid matrix
\begin{equation}
\Phi \in \mathbb R^{N_{\mathrm{basis}} \times N_{\mathrm{grid}}},
\qquad
\Phi_{\mu i} = \chi_\mu(\mathbf r_i),
\label{eq:phi_def}
\end{equation}
together with the diagonal grid-weight matrix
\begin{equation}
\mathbf{W} = \mathrm{diag}(w_i v_i^{\mathrm{xc}})
\in
\mathbb R^{N_{\mathrm{grid}} \times N_{\mathrm{grid}}}.
\label{eq:w_def}
\end{equation}
Let $\mathbf{D} \in \mathbb R^{N_{\mathrm{basis}} \times N_{\mathrm{basis}}}$ denote the density matrix. Then one of the dominant intermediates in EXC evaluation can be written as
\begin{equation}
\mathbf{R} = \mathbf{D}\Phi,
\label{eq:r_equals_dphi}
\end{equation}
and the EXC contribution itself becomes
\begin{equation}
\mathbf{F}^{\mathrm{xc}} = b\, \Phi \mathbf{W} \Phi^\top.
\label{eq:fxc_matrix_form}
\end{equation}
Equations~\eqref{eq:r_equals_dphi} and \eqref{eq:fxc_matrix_form} show that the dominant EXC workload can be expressed as a sequence of matrix products of the generic form
\begin{equation}
\mathbf{C} = \mathbf{A}\mathbf{B},
\label{eq:generic_matmul}
\end{equation}
which we take as the computational abstraction for the remainder of this work.

\subsection{Locality-Induced Structure}

Although the EXC workload can be expressed as matrix products, the resulting operands are not dense in practice. In Gaussian-basis electronic structure calculations, basis functions are spatially localized, so their values decay rapidly away from their associated atomic centers. As a result, for any given grid region, only a subset of basis functions contributes significantly to the numerical quadrature. This locality carries over to the matrix products in Eqs.~\eqref{eq:r_equals_dphi} and \eqref{eq:fxc_matrix_form}, where many entries are numerically negligible even though the matrices are formally dense.

Importantly, this sparsity is not random. Because retained basis contributions are induced by spatial proximity, significant entries tend to appear in clustered regions associated with nearby atoms, neighboring basis groups, or spatially compact grid partitions. Moreover, the size and location of these retained regions vary across grid batches: different batches may involve substantially different active basis supports depending on molecular geometry, pruning, and screening thresholds. The resulting structure is therefore both \emph{clustered} and \emph{dynamic}, rather than uniformly sparse or fixed throughout the calculation.

This regime is poorly matched to standard GPU execution strategies. Dense batched implementations regularize heterogeneous retained supports into common matrix shapes, which introduces padding, wasted memory traffic, and unnecessary arithmetic when different grid batches activate different subsets of basis functions. At the other extreme, generic fine-grained sparse formats incur substantial metadata overhead and irregular memory access, making it difficult to preserve arithmetic intensity and data locality on modern GPUs. EXC integration therefore falls into an intermediate regime: its dominant matrix products are neither efficiently dense nor well served by generic sparse kernels.

These observations motivate the design of KerneLDI. Rather than reverting to dense batching or element-wise sparse execution, KerneLDI preserves locality at block granularity, filters numerically negligible regions before multiplication, and exposes enough regular structure to support efficient GPU execution.

\subsection{Design Implications}

The locality-driven EXC workload imposes three requirements on an effective GPU implementation. First, the representation must preserve heterogeneous retained supports without forcing them into a common dense shape. Second, irregularity must be amortized at a coarser granularity than element-wise sparsity so that metadata and indirection do not dominate useful arithmetic. Third, the resulting computation must still expose enough regular structure to map efficiently onto GPU memory hierarchies, matrix units, and multi-GPU execution.

KerneLDI is designed around these requirements. It reorganizes EXC operands into fixed-size retained blocks, executes only compatible retained block pairs with dense block-level operators, and distributes heterogeneous grid-batch tasks dynamically across devices. The next section details this representation, execution model, and runtime design.

\section{KerneLDI}

\subsection{Overview and Design Goals}

Given the locality-driven and batch-dependent structure of the EXC matrix products described in Section~II, KerneLDI is designed to execute only numerically relevant interactions while avoiding both dense padding and fine-grained sparse indirection. The central idea is to preserve the clustered structure induced by spatial locality, but to do so at block granularity so that the resulting computation remains compatible with high-throughput GPU execution.

Figure~\ref{fig:2} summarizes the KerneLDI workflow. The framework proceeds in three stages. First, it transforms the original EXC operands into a block-filtered representation through locality-preserving reordering, fixed-size block partitioning, and block-level filtering. This stage converts formally dense but locality-structured matrices into a compact representation that retains only numerically significant blocks. Second, it executes the retained computation using dense block-level multipliers that operate on block pairs rather than on padded dense batches or element-wise sparse entries. Third, it extends the same representation and execution model to multi-GPU platforms by decomposing the filtered EXC workload into heterogeneous grid-batch tasks that are scheduled dynamically across devices.

These design choices directly target the requirements identified in Section~II. The block-filtered representation preserves locality without enforcing a common dense shape across heterogeneous grid batches; the block-level execution model amortizes sparse irregularity while recovering structured data reuse; and the task-based runtime improves scalability when the retained work varies substantially across batches. The remainder of this section details these three components.

\subsection{Block-Filtered Representation}

\subsubsection{Blockization and Notation}

KerneLDI operates on fixed-size blocks rather than on individual matrix entries. Let
\(
\mathbf{A} \in \mathbb R^{M \times K}
\)
and
\(
\mathbf{B} \in \mathbb R^{K \times N}
\)
denote a pair of operand matrices arising from the EXC formulations in Section~II, and let
\(
\mathbf{C} = \mathbf{A}\mathbf{B}
\)
be the corresponding output. We partition each operand into square blocks of size
\(
d \times d
\),
which serve as the fundamental unit of storage, filtering, and execution throughout the framework.

Under this partitioning, matrix
\(
\mathbf{A}
\)
is divided into
\(
M_r=\lceil M/d\rceil
\)
block rows and
\(
K_c=\lceil K/d\rceil
\)
block columns, yielding block entries
\(
\mathbf{A}_{pq}\in\mathbb R^{d\times d}
\)
for
\(
0 \le p < M_r
\)
and
\(
0 \le q < K_c
\).
Similarly,
\(
\mathbf{B}
\)
is divided into
\(
K_c
\)
block rows and
\(
N_c=\lceil N/d\rceil
\)
block columns, with block entries
\(
\mathbf{B}_{qr}\in\mathbb R^{d\times d}
\).
The output matrix
\(
\mathbf{C}
\)
is therefore partitioned into blocks
\(
\mathbf{C}_{pr}\in\mathbb R^{d\times d}
\),
where each output block is formed by accumulating products over compatible inner-dimension block indices:
\begin{equation}
\mathbf{C}_{pr} = \sum_q \mathbf{A}_{pq} \mathbf{B}_{qr}.
\label{eq:block_matmul}
\end{equation}

This blockization serves two purposes. First, it defines a coarser-grained execution unit that is large enough to amortize sparse metadata and indexing overhead. Second, it exposes computation in a form that can be mapped naturally to shared-memory staging and Tensor-Core-based dense block multiplication on GPUs. KerneLDI therefore does not attempt to preserve fine-grained sparsity explicitly; instead, it reorganizes the EXC workload into a block-structured form in which locality can be filtered and exploited at the level of retained matrix blocks.

\subsubsection{Locality-Preserving Reordering}

Block partitioning alone is not sufficient to produce an efficient representation, because the quality of the resulting blocks depends strongly on the ordering of rows and columns before partitioning. If spatially related entries are scattered in the original operand layout, fixed-size blocking will fragment numerically significant regions across many blocks and reduce the effectiveness of subsequent filtering. KerneLDI therefore reorders both grid points and basis functions prior to blockization so that spatially correlated entries become more compact in memory.

For grid points, KerneLDI applies a locality-preserving ordering based on atomic partitioning followed by Morton sorting within each atomic region. Grid points are first grouped according to their associated atoms, and the points within each group are then ordered by a Z-order space-filling curve. Given a grid point with discretized coordinates $(x,y,z)$, its Morton code is constructed by bit-interleaving the coordinate components into a one-dimensional key, which is then used for sorting. This procedure maps nearby three-dimensional coordinates to nearby one-dimensional positions while largely preserving spatial proximity, thereby reducing fragmentation when the basis-on-grid matrices are later partitioned into fixed-size blocks.

For basis functions, KerneLDI applies a similarity-based ordering derived from the overlap structure of the basis set. The key observation is that block partitioning is performed over matrix indices rather than over physical space. If basis indices with similar spatial support are scattered throughout the operand layout, then numerically significant contributions associated with a given grid region will be split across many blocks, weakening the effect of block filtering. KerneLDI therefore seeks an ordering in which basis functions with similar spatial interaction patterns appear close to one another in memory.

To this end, KerneLDI uses the overlap matrix
\begin{equation}
\mathbf{S}_{ij}=\int \phi_i(\mathbf r)\phi_j(\mathbf r)\, d\mathbf r
\label{eq:overlap_matrix}
\end{equation}
as a practical proxy for spatial and chemical locality. For each basis function $i$, we define an overlap signature
\begin{equation}
\mathbf s_i = (\mathbf{S}_{i1}, \mathbf{S}_{i2}, \dots, \mathbf{S}_{iN_{\mathrm{basis}}}),
\label{eq:overlap_signature}
\end{equation}
and measure the similarity between basis functions $i$ and $j$ using cosine similarity,
\begin{equation}
\mathrm{sim}(i,j)=
\frac{\mathbf s_i \cdot \mathbf s_j}
{\|\mathbf s_i\|\, \|\mathbf s_j\|}.
\label{eq:cosine_similarity}
\end{equation}
Basis functions with similar overlap signatures tend to occupy nearby spatial regions or share similar local chemical environments, and are therefore more likely to be activated together on the same grid batches. KerneLDI places such basis functions close to one another in the reordered layout using similarity-based clustering, so that the active basis support associated with a grid region is concentrated into a smaller number of contiguous index ranges.

The purpose of this reordering step is not merely to permute the operands, but to improve block compactness before filtering. After reordering, numerically significant entries are concentrated into fewer and denser blocks, boundary fragmentation is reduced, and the retained block pattern becomes more amenable to efficient block-level execution on GPUs. Implementation details of the Morton ordering, overlap-signature construction, and basis-function clustering are provided in Appendix~\ref{sec:reordering_details}.

\subsubsection{Block-Level Filtering}

After reordering and block partitioning, KerneLDI applies block-level filtering to discard numerically negligible regions before multiplication. The goal of this step is to reduce both storage and computation by retaining only those blocks whose contributions are potentially significant, while avoiding the metadata and access overhead of fine-grained sparse representations.

For each block \(A_{pq}\), KerneLDI computes a block summary based on its maximum absolute entry,
\begin{equation}
a_{pq} = \|\mathbf{A}_{pq}\|_{\max}
= \max_{u,v} |(\mathbf{A}_{pq})_{uv}|,
\label{eq:block_summary_A}
\end{equation}
and retains the block only if
\begin{equation}
a_{pq} \ge t_b,
\label{eq:block_filter_A}
\end{equation}
where \(t_b\) is the block-retention threshold. The retained block set of \(\mathbf{A}\) is therefore
\begin{equation}
\mathcal S_A
=
\{(p,q)\mid \|\mathbf{A}_{pq}\|_{\max} \ge t_b\}.
\label{eq:retained_blocks_A}
\end{equation}
The same procedure is applied to \(\mathbf{B}\), yielding
\begin{equation}
\mathcal S_B
=
\{(q,r)\mid \|\mathbf{B}_{qr}\|_{\max} \ge t_b\}.
\label{eq:retained_blocks_B}
\end{equation}

This filtering step serves two purposes. First, it removes blocks whose entries are uniformly too small to contribute meaningfully to the subsequent multiplication. Second, it shrinks the candidate search space for block-pair execution, since only products involving retained blocks need to be considered downstream. Because the preceding reordering concentrates significant entries into spatially compact regions, the retained blocks after filtering are typically denser and more clustered than they would be under the original ordering.

Importantly, block-level filtering does not yet determine which block pairs are multiplied. It only defines a compact operand representation by removing negligible blocks independently in each matrix. The finer-grained selection of executable block pairs is performed during multiplication, as described in the next subsection.

In the present implementation, reordering and block-level filtering are performed once for a given molecular geometry and the resulting block structure is reused across repeated EXC evaluations. Under a fixed geometry, the locality pattern that determines block retention remains unchanged, while only the numerical values associated with the retained blocks are updated. This design amortizes the preprocessing overhead rather than incurring it at every evaluation.

\subsubsection{Dual Block-Compressed Layouts}

After block-level filtering, KerneLDI stores the retained blocks in block-compressed layouts tailored to the role of each operand in multiplication. The left operand \(\mathbf{A}\) is stored in a row-oriented block-compressed format, denoted BCS(R), while the right operand \(\mathbf{B}\) is stored in a column-oriented format, denoted BCS(C). This dual-layout design is chosen to make the retained block intersections that contribute to each output block directly enumerable during multiplication.

For the left operand, BCS(R) groups all retained blocks that share the same block row. Concretely, for each block row \(p\), the layout stores the list of retained column indices \(q\) such that \((p,q)\in\mathcal S_A\), together with the corresponding dense block data \(\mathbf{A}_{pq}\). Symmetrically, for the right operand, BCS(C) groups all retained blocks that share the same block column. For each block column \(r\), the layout stores the list of retained row indices \(q\) such that \((q,r)\in\mathcal S_B\), together with the corresponding dense block data \(\mathbf{B}_{qr}\). In practice, different EXC matrix products may involve non-transposed or transposed operands (e.g., NN or TN forms), and KerneLDI chooses the corresponding block-compressed orientation so that the retained inner-dimension intersections can still be traversed efficiently.

This organization is important for efficient multiplication. For a given output block \(\mathbf{C}_{pr}\), KerneLDI needs to identify the compatible inner-dimension block indices \(q\) for which both \(\mathbf{A}_{pq}\) and \(\mathbf{B}_{qr}\) are retained. Under the dual block-compressed layouts, this becomes the intersection
\begin{equation}
\mathcal I(p,r)
=
\{\, q \mid (p,q)\in\mathcal S_A,\ (q,r)\in\mathcal S_B \,\},
\label{eq:block_intersection}
\end{equation}
and the output block can be written as
\begin{equation}
\mathbf{C}_{pr}
=
\sum_{q\in\mathcal I(p,r)} \mathbf{A}_{pq} \mathbf{B}_{qr}.
\label{eq:block_output_intersection}
\end{equation}
The dual layouts therefore align the operand representation with the block-pair execution pattern, allowing KerneLDI to enumerate only those retained block pairs that can contribute to a given output block.

Importantly, this design is not intended merely to compress storage. Its main purpose is to preserve the locality exposed by filtering while supporting efficient traversal of retained block pairs during GPU execution. By storing left operands contiguously by block row and right operands contiguously by block column, KerneLDI reduces indexing overhead, improves memory locality when forming block intersections, and provides the direct input structure for the executable block-pair selection described next.

\subsection{Block-Pair Execution Operator}

\subsubsection{Executable Pair Selection}

Block-level filtering produces compact operand representations, but it does not by itself determine which retained block pairs should be multiplied. For a given output block \(\mathbf{C}_{pr}\), KerneLDI only needs to consider inner-dimension block indices \(q\) that lie in the retained intersection set \(\mathcal I(p,r)\) defined in Eq.~\eqref{eq:block_intersection}. Even within this intersection, however, some retained block pairs may still be numerically negligible and need not be executed.

KerneLDI therefore applies a second filtering stage at the level of block pairs. For each retained block \(\mathbf{A}_{pq}\) and \(\mathbf{B}_{qr}\), we reuse the block summaries introduced in Eqs.~\eqref{eq:block_summary_A}--\eqref{eq:retained_blocks_B} and form the pairwise screening criterion
\begin{equation}
a_{pq}\, b_{qr} \ge t_d,
\label{eq:block_pair_filter}
\end{equation}
where \(a_{pq}=\|\mathbf{A}_{pq}\|_{\max}\), \(b_{qr}=\|\mathbf{B}_{qr}\|_{\max}\), and \(t_d\) is the block-pair execution threshold. Only those retained block pairs that satisfy Eq.~\eqref{eq:block_pair_filter} are dispatched to the multiplication kernel.

This two-level filtering serves distinct purposes. The block-retention threshold \(t_b\) removes negligible blocks independently in each operand and defines the compact block-filtered representation. The block-pair threshold \(t_d\), in contrast, is applied only after compatible retained intersections have been identified, and further reduces the executed work by screening out block pairs whose joint contribution is too small to justify multiplication. As a result, the set of executed block pairs can be substantially smaller than the Cartesian product of retained blocks in the two operands.

For each output block \(\mathbf{C}_{pr}\), the executed computation is therefore
\begin{equation}
\mathbf{C}_{pr}
=
\sum_{q \in \mathcal E(p,r)} \mathbf{A}_{pq} \mathbf{B}_{qr},
\label{eq:executed_block_output}
\end{equation}
where
\begin{equation}
\mathcal E(p,r)
=
\{\, q \in \mathcal I(p,r) \mid a_{pq}\, b_{qr} \ge t_d \,\}
\label{eq:executed_block_pairs}
\end{equation}
denotes the set of executable retained block pairs. This executable-pair formulation provides the immediate interface between the block-filtered representation and the GPU multiplication kernel described next.

\subsubsection{GPU Kernel Mapping}

Once the executable block pairs have been identified, KerneLDI evaluates them using a dense block-level multiplication kernel specialized to the filtered workload. The key design choice is to treat each retained block pair as a unit of dense computation rather than reverting either to padded dense batches or to element-wise sparse execution. This allows the kernel to preserve the arithmetic regularity of dense matrix multiplication while restricting work to numerically relevant regions only.

For each executable pair \((\mathbf{A}_{pq}, \mathbf{B}_{qr})\), the corresponding dense \(d\times d\) blocks are loaded from the block-compressed layouts into shared memory buffers. GPU threads then perform block-level multiply--accumulate operations on these shared-memory tiles, with the resulting partial products accumulated into the corresponding output block \(\mathbf{C}_{pr}\). In the current implementation, these block multiplications are mapped to warp-level matrix operations through the WMMA interface, allowing the retained computation to benefit from Tensor Core acceleration while maintaining the block-filtered execution model. For fused evaluations of the electron density and its derivatives, KerneLDI additionally exploits symmetry in the underlying AO-space contraction to reduce redundant block-pair computation and accumulation. Retained pairs are compactly enumerated via parallel prefix-sum compaction (Appendix~\ref{sec:para_add}) and then assigned to thread blocks in block-pair units, while partial products targeting the same output tile are accumulated before global write-back.

This execution strategy is designed to recover the main performance advantages of dense GPU kernels at block granularity. Shared-memory staging reduces repeated global-memory traffic for retained blocks, block-level multiply--accumulate exposes regular data reuse, and warp-level dense execution avoids the control divergence and irregular access patterns typical of fine-grained sparse kernels. At the same time, because only executable retained pairs are processed, the kernel cost scales with the filtered block-pair workload rather than with the full dense product space.

In practice, multiple executable pairs may contribute to the same output block \(\mathbf{C}_{pr}\). KerneLDI therefore organizes the kernel around block-pair accumulation: each executed pair produces a partial \(d\times d\) contribution, and these partial results are accumulated into the corresponding output block before write-back. The precise thread-block mapping and accumulation strategy are implementation dependent, but the governing abstraction remains the same: dense computation is performed only on those block pairs that survive locality-aware filtering.

\subsubsection{Why This Regime Favors Block-Dense Execution}

The execution regime targeted by KerneLDI is not well served by either standard dense batching or generic sparse kernels. Dense batched execution requires heterogeneous retained supports to be regularized into common shapes, which introduces padding and unnecessary arithmetic when different grid batches activate different subsets of basis functions. Generic sparse kernels avoid this dense overhead, but they typically operate at too fine a granularity to make effective use of GPU shared memory, warp-level matrix units, and structured data reuse.

KerneLDI therefore adopts an intermediate strategy: it deliberately gives up fine-grained sparsity in order to recover dense execution regularity at the level of retained blocks. This tradeoff is appropriate because the EXC workload is neither fully dense nor extremely sparse; its significant contributions are clustered by locality and thus can be compacted into dense block tiles after reordering and filtering. By operating on these retained tiles, KerneLDI amortizes sparse indexing overhead over substantially more useful arithmetic than element-wise sparse execution would allow.

This block-dense formulation yields three practical advantages. First, it aligns naturally with GPU memory hierarchies, since retained blocks can be staged contiguously and reused within shared memory. Second, it exposes compute tiles that are compatible with high-throughput dense matrix instructions, including Tensor Core operations. Third, it reduces the amount of wasted work relative to dense batching because only retained and executable block pairs are processed. For locality-driven EXC matrix products, this combination makes block-dense execution a better match to the workload structure than either padded dense kernels or generic sparse formats.

\subsection{Complexity and Cost Model}

The cost of KerneLDI is governed by the size of the filtered block workload rather than by the full dense product space. For a given matrix product \(\mathbf{C}=\mathbf{AB}\), let \(n_A = |\mathcal S_A|\) and \(n_B = |\mathcal S_B|\) denote the numbers of retained blocks in the two operands after block-level filtering, and let \(n_P\) denote the total number of executable block pairs that survive the pairwise screening in Eq.~\eqref{eq:block_pair_filter}. Under this notation, the cost of KerneLDI can be separated into two components: a preprocessing stage and an execution stage. The preprocessing stage includes reordering, block partitioning, block summary construction, filtering, and block-compressed layout generation. The execution stage consists of dense block multiplications over the \(n_P\) executable block pairs together with the associated block-level accumulation into output tiles. Consequently, the dominant arithmetic cost is governed by the filtered workload size and scales approximately with \(n_P \, d^3\), rather than with the full dense product space alone.

This separation explains the performance regime targeted by KerneLDI. On smaller systems, or in cases where locality is not yet strong enough to eliminate a substantial fraction of the block pairs, the one-time preprocessing overhead can account for a noticeable portion of the total runtime. As system size increases, however, the spatial locality of the EXC workload becomes more pronounced, the fraction of retained and executable block pairs decreases relative to the dense block space, and the reduction in multiplication work increasingly outweighs the preprocessing cost. In the present implementation, the block structure is constructed once for a fixed molecular geometry and reused across repeated EXC evaluations, so the preprocessing overhead is amortized rather than paid at every evaluation. KerneLDI therefore exhibits a characteristic crossover behavior: the gains are moderate when the filtered workload is still close to dense, but become substantially larger once locality-induced block sparsity dominates the computation.

\subsection{Multi-GPU Task Decomposition and Dynamic Scheduling}

KerneLDI extends naturally to multi-GPU execution by decomposing the filtered EXC workload into grid-batch tasks. Each task corresponds to a filtered grid batch together with its associated retained basis support and block-pair workload, and produces a partial contribution to the EXC matrices or related quantities. Because the retained structure is determined by locality and screening, different grid batches generally contain different numbers of retained blocks and executable block pairs. The resulting task costs are therefore heterogeneous even within the same calculation, making uniform static partitioning ineffective once the workload is distributed across multiple devices.

To address this heterogeneity, KerneLDI adopts a pull-based dynamic scheduling strategy. Filtered grid-batch tasks are maintained in a shared task pool, and GPU workers acquire new tasks on demand as soon as they complete their current work. Each GPU evaluates its assigned tasks independently using the same block-filtered representation and block-pair execution operator described above, producing partial EXC contributions locally. After all tasks have been processed, the partial results are combined through a final reduction step to recover the complete EXC output. This runtime design improves load balance without changing the underlying numerical formulation: communication is confined to task coordination and final reduction, while the bulk of the filtered block computation remains device-local.

\subsection{Numerical Considerations and Scope}

The numerical approximation introduced by KerneLDI is controlled primarily by the block-retention threshold \(t_b\) and the block-pair execution threshold \(t_d\), which determine which operand blocks and block products are discarded prior to multiplication. Numerical errors for both energies and gradients are evaluated empirically in Section~IV. The current implementation targets local and semilocal EXC functionals, where the dominant computation admits the locality-driven matrix-product form described in Section~II; extensions to nonlocal correlation terms remain future work.

\begin{figure}[t!]
    \centering
    \includegraphics[width=0.85\linewidth]{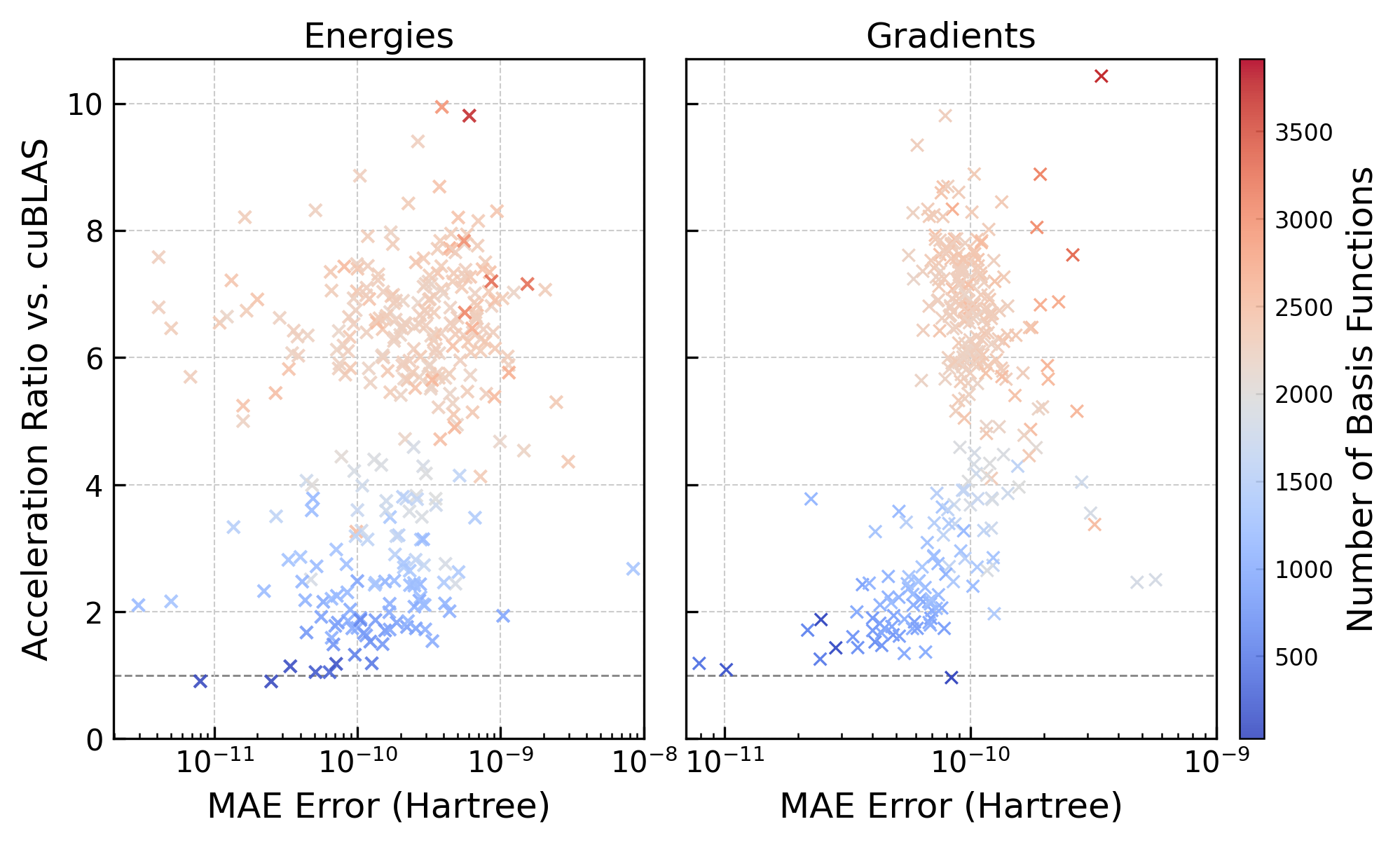}
    \caption{Single-GPU speedup relative to dense GPU execution. Each point denotes one molecule from the dataset~\cite{ju2024acceleration} and is colored by basis size. The horizontal axis reports the mean absolute error in energies (left) or gradients (right), while the vertical axis reports KerneLDI speedup over EXC-cuBLAS.}
    \label{fig:4}
\end{figure}

\section{Results}

We evaluate KerneLDI along five axes: correctness and sensitivity of the EXC approximation, scaling performance with increasing system size, end-to-end SCF evaluation, multi-GPU scaling, and impact on \textit{ab initio} molecular dynamics.

\paragraph{Baselines.}
For standalone EXC component benchmarks, the baselines are a CPU dense implementation using OpenBLAS (EXC-OpenBLAS) and a GPU dense implementation using cuBLAS (EXC-cuBLAS), with all remaining electronic-structure components following MADFT~\cite{ju2024acceleration}. For the end-to-end SCF evaluation (Section~IV-C), we additionally compare against GPU4PySCF~\cite{wu2025enhancing}, a widely used GPU-accelerated DFT implementation based on dense batched execution.

\paragraph{Hardware platform.}
Single-GPU experiments use one NVIDIA Tesla V100-PCIE-16GB (5120 CUDA cores, 640 Tensor Cores, 16\,GB HBM2, 900\,GB/s memory bandwidth) driven by CUDA Toolkit 12.2 and cuBLAS 12.2.
CPU baseline runs use an AMD EPYC 7452 (32 cores, 2.35\,GHz base / 3.35\,GHz boost, 128\,MB L3 cache) with OpenBLAS 0.3.21.
Multi-GPU scaling experiments are conducted on nodes equipped with multiple V100 GPUs connected by NVLink, with inter-node communication over InfiniBand and coordination through OpenMPI.

\paragraph{Numerical settings.}
The block size is $d=32$, and both filtering thresholds are set to $t_b=t_d=10^{-12}$ in all experiments. EXC numerical integration uses a grid equivalent to the PySCF level-4 setting. The ERI screening tolerance is $\epsilon_{\text{ERI}}=10^{-12}$\,Hartree and the SCF energy convergence threshold is $10^{-8}$\,Hartree. All experiments use the M06-2X functional~\cite{zhao2008m06}. The end-to-end SCF (Section~IV-C) and multi-GPU scaling (Section~IV-D) experiments use the def2-TZVP basis set~\cite{weigend2005balanced}; all other experiments use the def2-SVP basis set~\cite{weigend2005balanced}.
All reported timing results are averaged over three independent runs.

\begin{figure}[t]
    \centering
    \includegraphics[width=0.85\columnwidth]{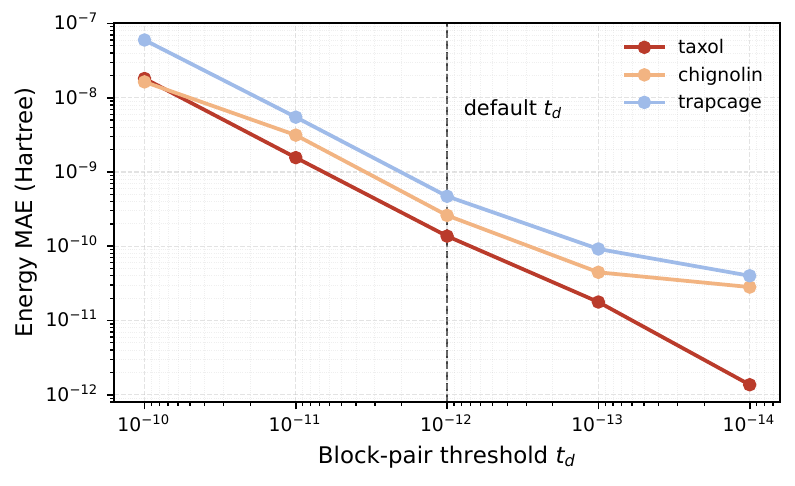}
    \caption{Sensitivity of EXC energy accuracy to the block-pair threshold \(t_d\). The vertical dashed line marks the default setting used in the main experiments. Error remains low near the default threshold and increases systematically as \(t_d\) is increased.}
    \label{fig:td_sensitivity}
\end{figure}

\subsection{Correctness, Threshold Sensitivity, and Single-GPU Performance}

To assess correctness and single-GPU throughput across chemically diverse workloads, we compare KerneLDI against the dense cuBLAS baseline on the 329-molecule dataset proposed in~\cite{ju2024acceleration}. The dataset contains large molecular systems, including transition-metal complexes, spanning elements from the first through fourth rows of the periodic table; each molecule contains at least 100 atoms and at least four distinct atom types.

Figure~\ref{fig:4} shows that KerneLDI preserves numerical accuracy while delivering substantial acceleration over dense GPU execution. For both energies and gradients, the mean absolute error remains in the range of roughly \(10^{-11}\) to \(10^{-9}\) Hartree, while the speedup increases with system size and reaches up to about one order of magnitude on the largest systems. This trend is consistent with the locality-driven sparsity characteristics described in Section~II: as the basis and grid grow, the retained structure becomes more exploitable and the cost of padded dense batching becomes increasingly pronounced.

To further examine the approximation introduced by pair filtering, we sweep the block-pair execution threshold \(t_d\) on three representative systems while keeping all other settings fixed. Figure~\ref{fig:td_sensitivity} reports the resulting mean absolute error (MAE) in EXC energies for taxol, chignolin, and trapcage, complementing the dataset-level accuracy results above by showing how the error varies around the default operating point.

Across all three systems, the energy error remains very small in the vicinity of the default threshold \(t_d = 10^{-12}\). As \(t_d\) is increased, the screening criterion in Eq.~\eqref{eq:block_pair_filter} becomes more selective, fewer block pairs are executed, and the resulting approximation error increases. The trend is monotonic for all three systems, and the default setting remains in the low-error regime in each case. Together with the dataset-level accuracy results in Fig.~\ref{fig:4}, this shows that block-pair filtering remains well controlled in the operating regime used throughout the paper.

\subsection{Scaling Performance of KerneLDI}


To examine how the method behaves as sparsity becomes more pronounced, we evaluate six representative molecules whose atom counts increase progressively: porphy (85 atoms), taxol (110 atoms), chignolin (166 atoms), trapcage (272 atoms), olestra (453 atoms), and crambin (642 atoms). This setup is intended to expose the crossover from regimes where sparse data structures provide limited benefit to regimes where locality-driven sparsity dominates the computation.

\begin{figure}[t!]
    \centering
    \begin{subfigure}[b]{\linewidth}
        \centering
        \includegraphics[width=\linewidth]{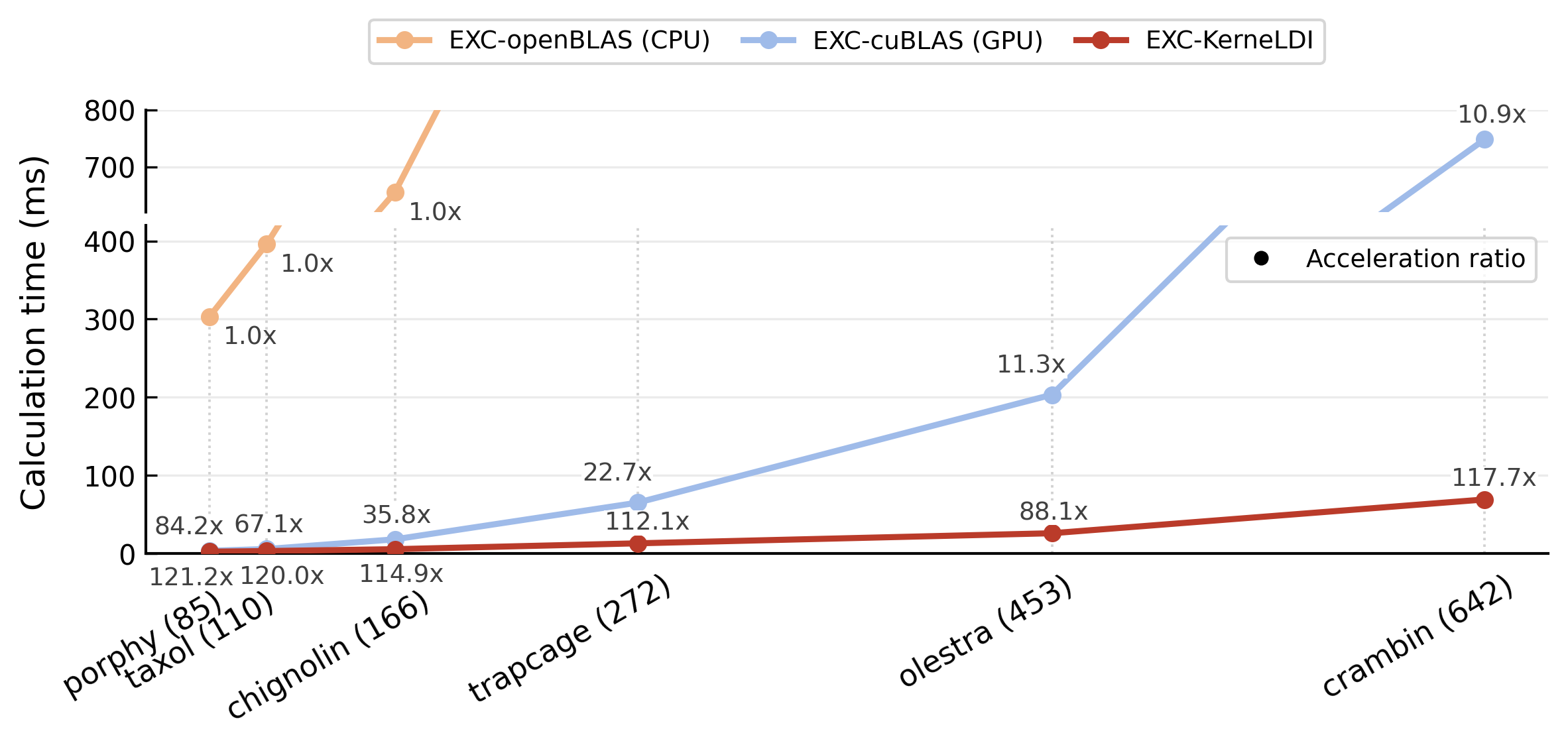} 
        \caption{EXC energy runtime across six molecular systems.\\}
        \label{fig:subfig1}
    \end{subfigure}
    \hfill
    \begin{subfigure}[b]{\linewidth}
        \centering
        \includegraphics[width=\linewidth]{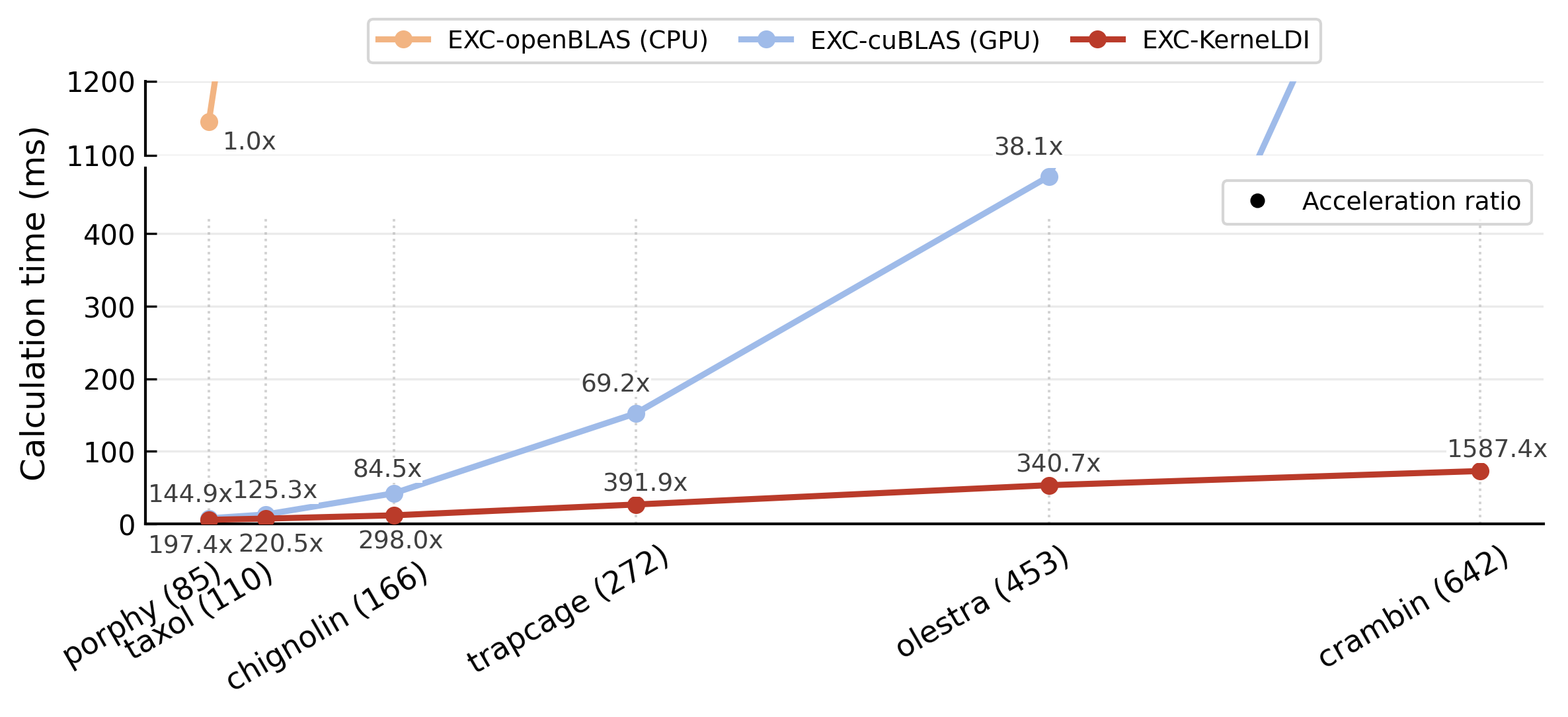} 
        \caption{EXC gradient runtime for the same systems.}
        \label{fig:subfig2}
    \end{subfigure}
    \caption{Crossover behavior of KerneLDI under increasing system size. Each panel shows calculation time (ms) on the vertical axis for three methods plotted against six molecular systems of growing size on the horizontal axis (porphy (85) through crambin (642), where the trailing number denotes the number of atoms). The annotated × markers indicate the acceleration ratio of KerneLDI relative to EXC-openBLAS at each system size.}
    \label{fig:5}
\end{figure}

Figures~\ref{fig:subfig1} and \ref{fig:subfig2} reveal a clear crossover behavior. The dense openBLAS and cuBLAS baselines continue to grow much more rapidly with system size because they execute dense matrix operations regardless of how much of the underlying EXC structure is numerically negligible. KerneLDI, by contrast, increasingly benefits from screening and block-structured execution as the systems become larger and more weakly coupled. The widening gap shows that the method is best matched to the large-system regime where padded dense batching becomes least efficient.

For the smaller molecules, the overhead of block construction, filtering, and sparse-aware execution partially offsets the reduction in arithmetic work, so the performance gap remains moderate. Once the workload reaches roughly 200--300 atoms, corresponding to about 1500--2000 basis functions in this setup, the reduction in screened multiplication work becomes the dominant factor in the overall speedup. Beyond that crossover point, the KerneLDI curves grow much more slowly than the dense baselines for both energy and gradient evaluation. The key point is therefore not simply that KerneLDI is faster, but that its advantage strengthens exactly in the large-system regime where dense batching becomes least well matched to the workload.

\subsection{End-to-End SCF Performance Evaluation}
\label{sec:e2e_performance}

\begin{figure}[t]
    \centering
    \includegraphics[width=\linewidth]{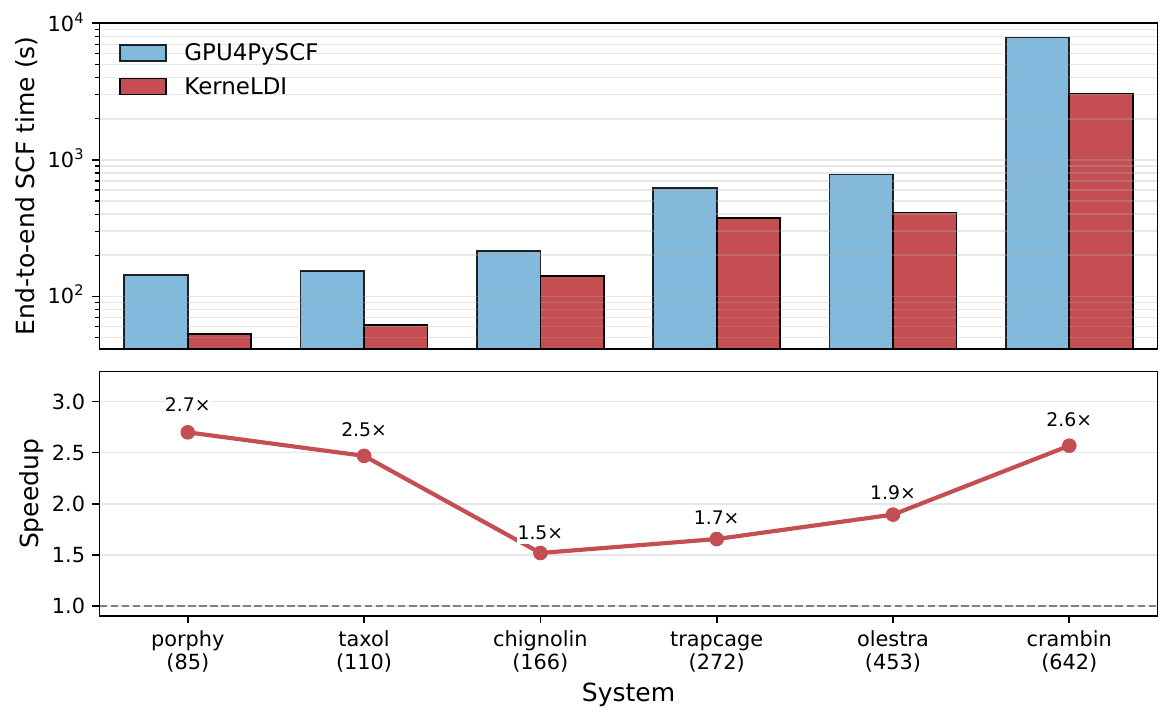}
    \caption{End-to-end SCF comparison across six molecular systems.
    The top panel reports the total SCF wall-clock time for GPU4PySCF and
    KerneLDI, and the bottom panel reports the corresponding speedup of
    KerneLDI over GPU4PySCF.
    }
    \label{fig:scf_e2e}
\end{figure}

To determine how the kernel-level acceleration translates into application-level
benefit, we next evaluate full self-consistent field (SCF) calculations.
As a practical baseline, we compare against GPU4PySCF, a widely used
GPU-accelerated implementation based on dense batched execution.
All comparisons are performed on the same hardware under identical numerical
settings and convergence criteria.

Figure~\ref{fig:scf_e2e} shows that KerneLDI consistently improves end-to-end
SCF performance across all six test systems.
The observed speedups over GPU4PySCF range from 1.52$\times$ to 2.70$\times$,
demonstrating that the advantages of the block-filtered EXC execution model are
preserved at the level of complete SCF calculations rather than being confined
to an isolated kernel benchmark.

For the individual systems, the speedup is 2.70$\times$ for porphy,
2.47$\times$ for taxol, 1.52$\times$ for chignolin, 1.65$\times$ for trapcage,
1.89$\times$ for olestra, and 2.57$\times$ for crambin.
This consistency is important because it shows that the advantage of KerneLDI
is retained at the level of complete SCF calculations rather than appearing
only in an isolated EXC kernel benchmark.
These results show that preserving locality-driven structure in the EXC pathway
yields a clear end-to-end performance advantage in complete SCF calculations.

Overall, these results indicate that KerneLDI improves the throughput of full
DFT calculations in a practically meaningful way.
Rather than only accelerating a standalone EXC kernel, the proposed
representation and execution model reduces the time-to-solution of complete SCF
workflows across a chemically diverse set of systems.

\subsection{Multi-GPU Scaling Performance}

\begin{figure}[t]
    \centering
    \includegraphics[width=\linewidth]{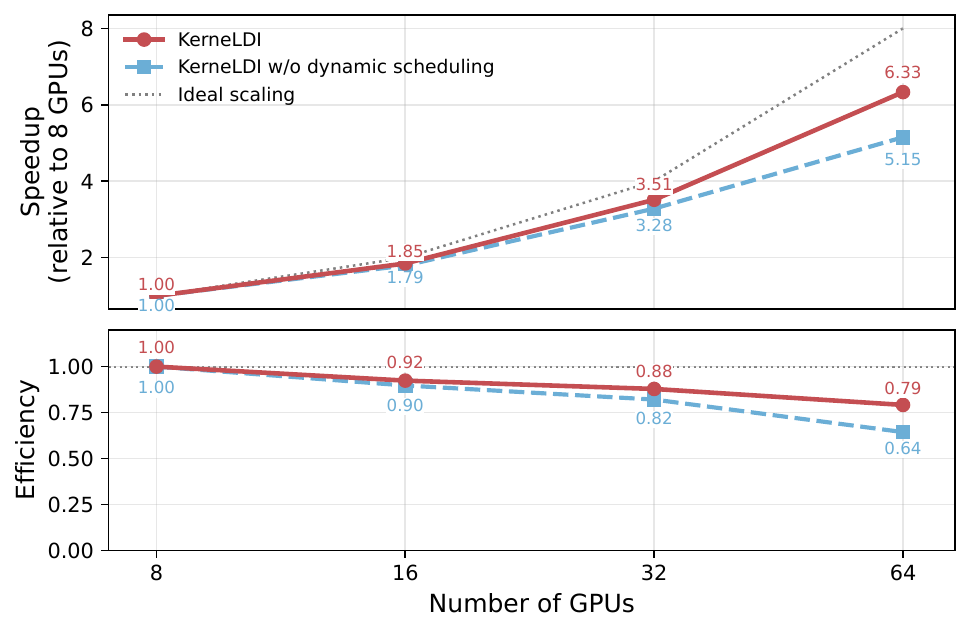}
    \caption{Multi-GPU scaling of KerneLDI on ubiquitin from 8 to 64 GPUs.
    The top panel reports the speedup normalized to the 8-GPU configuration,
    and the bottom panel reports the corresponding parallel efficiency.
    We compare the full KerneLDI implementation against an ablated version
    without dynamic scheduling.
    }
    \label{fig:multi_gpu_scaling}
\end{figure}

To evaluate the scalability of KerneLDI on larger accelerator configurations,
we perform multi-GPU experiments on ubiquitin using 8, 16, 32, and 64 GPUs.
All runs use the same molecular system, basis set, integration settings, and
screening thresholds, so that the measured differences reflect only the effect
of GPU count and scheduling strategy.
We use the 8-GPU configuration as the reference point for normalized speedup
and parallel efficiency.

Figure~\ref{fig:multi_gpu_scaling} reports the resulting scaling behavior.
The full KerneLDI implementation exhibits strong scaling across the entire
8--64 GPU range, reaching speedups of 1.85$\times$, 3.51$\times$, and
6.33$\times$ at 16, 32, and 64 GPUs, respectively, relative to the 8-GPU
baseline.
The corresponding parallel efficiencies are 92.4\%, 87.8\%, and 79.2\%.
These results indicate that the filtered EXC workload decomposition of
KerneLDI exposes substantial task parallelism and maintains high utilization
as the number of GPUs increases.

To assess the contribution of the scheduler, we compare against an ablated
version of KerneLDI without dynamic scheduling.
This variant reaches speedups of 1.80$\times$, 3.28$\times$, and
5.15$\times$ at 16, 32, and 64 GPUs, with corresponding parallel efficiencies
of 89.7\%, 82.1\%, and 64.4\%.
Although the ablated version still scales, its efficiency drops more rapidly as
the GPU count increases.
The difference is modest at 16 GPUs but becomes increasingly pronounced at
32 and 64 GPUs, where the full KerneLDI implementation maintains noticeably
better scaling.

This behavior is consistent with the workload characteristics of
locality-driven integration.
After block filtering, different grid batches retain different numbers of basis
blocks and block-pair interactions, so their computational costs are
heterogeneous.
Without dynamic scheduling, this heterogeneity leads to increasing load
imbalance as the workload is distributed across more GPUs.
By assigning tasks on demand, KerneLDI reduces this imbalance and preserves
higher parallel efficiency at larger scale.

Overall, the ubiquitin results show that KerneLDI not only accelerates the EXC
workload on a single GPU configuration but also scales efficiently across a
substantial multi-GPU range.
The persistent gap between the full implementation and the version without
dynamic scheduling shows that the scheduling strategy is an important part of
making the block-filtered execution model effective at scale.

\subsection{Accelerating Molecular Dynamics with KerneLDI}

To illustrate the downstream consequence of faster EXC evaluation, we next examine ab initio molecular dynamics for AceAla15Lys, a capped polypeptide comprising 15 alanine residues followed by a C-terminal lysine. This system is a standard folding-oriented test case widely used in vacuum simulations. The initial structure is an extended linear chain. All simulations share the same initial configuration and identical computational settings; the only difference is the EXC evaluation method.

Rather than fixing the simulated physical time, we compare different
methods under the same wall-clock runtime. Since each MD step requires
a full electronic structure evaluation, faster EXC computation enables
a larger number of MD steps to be completed within the same runtime.
Consequently, the simulated physical time (measured in picoseconds) that
can be reached differs across methods.

\begin{figure}[t!]
    \centering
    \includegraphics[width=0.9\linewidth]{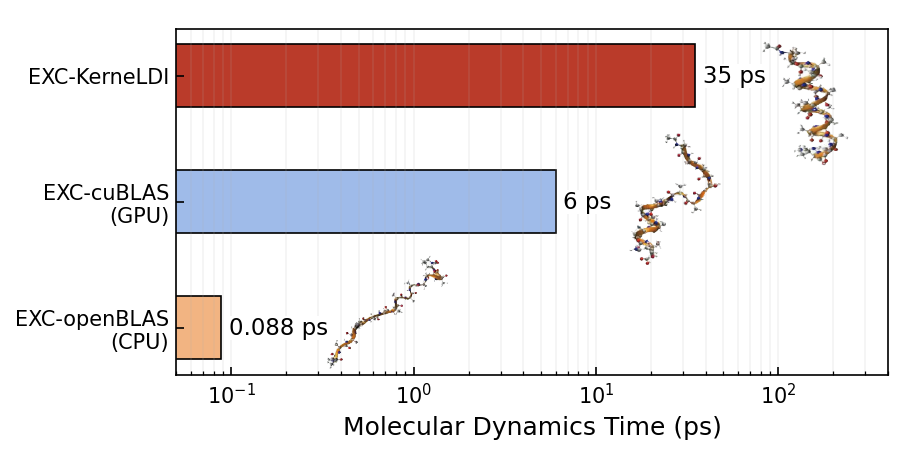}
    \caption{Simulated AIMD trajectory length achievable within a fixed wall-clock budget of 800 hours across three methods. Annotated values indicate the total simulated time achieved by each method within the same computational budget. The molecular snapshots of the AceAla15Lys system alongside each bar illustrate the corresponding structural evolution.}
    \label{fig:md_time}
\end{figure}

Starting from the same initial configuration and using identical
simulation parameters, the CPU baseline EXC-openBLAS is able to
propagate the trajectory by only 0.088~ps within the fixed runtime.
The GPU baseline EXC-cuBLAS implementation significantly improves the throughput,
reaching approximately 6~ps of simulated time under the same conditions.
With KerneLDI, the substantially reduced EXC computational cost allows
the simulation to advance to about 35~ps within the identical runtime.

The atomic configurations obtained at the end of these runs therefore
correspond to different stages of the dynamical evolution. The final
structures reached by the three approaches are shown in
Fig.~\ref{fig:md_time}, demonstrating that the improved EXC
performance of KerneLDI enables AIMD simulations to explore a much
longer physical timescale within the same computational budget. From a
systems perspective, this result is important because it converts kernel-level acceleration into directly usable scientific throughput. KerneLDI increases the amount of first-principles dynamics that can be simulated within a fixed wall-clock budget.


\section{Related Work}

\paragraph{Structured and block-sparse matrix computation on GPUs.}
Generic sparse libraries such as cuSPARSE~\cite{cusparse} provide element-wise formats (CSR, COO) optimized for highly sparse regimes, while block-sparse representations have been explored in sparse tensor processing~\cite{qin2021extending} and density-matrix methods~\cite{artemov2021sparse}. Hardware-level structured sparsity~\cite{choquette2021nvidia} doubles throughput but requires a fixed fine-grained pattern. KerneLDI addresses the intermediate regime where block occupancy varies dynamically and the sparsity pattern is determined at runtime by spatial locality.

\paragraph{GPU-accelerated numerical integration in DFT.}
GPU acceleration of EXC integration has been demonstrated via batched dense GEMM in GauXC~\cite{williams2021achieving}, QUICK~\cite{manathunga2020parallel}, and more recent pipelines~\cite{wu2025enhancing,stocks2025efficient}; Tensor Core acceleration has also been applied in this context~\cite{barca2020recent,markidis2018nvidia,haidar2018harnessing}. All of these approaches regularize EXC into uniformly shaped dense workloads. KerneLDI takes a different path by preserving and exploiting the block-level sparsity that batched dense strategies discard.

\paragraph{Dynamic task scheduling for irregular GPU workloads.}
Task-based scheduling and work-stealing strategies have been proposed for irregular GPU workloads in sparse linear algebra~\cite{anzt2022ginkgo}, molecular dynamics~\cite{gotz2012routine}, and general GPU task management~\cite{steinberger2014softshell}. KerneLDI adopts a pull-based dynamic scheduling strategy that assigns filtered grid batches to GPU devices on demand, addressing the load imbalance inherent in locality-driven integration without requiring a priori cost estimates.

\section{Future Work}

At present, KerneLDI is implemented for local and semi-local exchange--correlation functionals. Extensions to non-local correlation terms remain an important direction for future work. More broadly, because the block-filtered representation and dense block multiplier are designed around the general structure of locality-driven matrix multiplications rather than EXC-specific properties, we expect the framework to be applicable to other quantum-chemistry workloads that share this computational pattern, including Coulomb density fitting and Schwarz-screened exact-exchange construction.



\section{Acknowledgment}
This project was supported by the Zhongguancun Academy under the Internal Research Grant No. C20250501. Part of the early-stage work was conducted by X.W., Y.P., and F.J.\ during internship or affiliation at Microsoft Research. We sincerely thank Derk Pieter Kooi, David Williams-Young and Hongbin Liu for their helpful suggestions during the preparation of the initial version of this manuscripts, and Paola Gori Giorgi, Sebastian Ehlert, Jan Hermann and Zun Wang for their valuable insights into density functional theory and related computational software.






\bibliographystyle{bst/IEEEtran} 
\bibliography{bst/main}

@article{kohnsham1965self,
  title={Self-Consistent Equations Including Exchange and Correlation Effects},
  author={Kohn, Walter and Sham, Lu Jeu},
  journal={Physical Review},
  volume={140},
  number={4A},
  pages={A1133--A1138},
  year={1965},
  doi={10.1103/PhysRev.140.A1133},
  url={https://doi.org/10.1103/PhysRev.140.A1133}
}

@book{marx2009ab,
  title={Ab Initio Molecular Dynamics: Basic Theory and Advanced Methods},
  author={Marx, Dominik and Hutter, J{\"u}rg},
  publisher={Cambridge University Press},
  year={2009},
  isbn={9780521898638}
}

@article{kresse1996efficient,
  title={Efficient iterative schemes for ab initio total-energy calculations using a plane-wave basis set},
  author={Kresse, Georg and Furthm{\"u}ller, J{\"u}rgen},
  journal={Physical Review B},
  volume={54},
  number={16},
  pages={11169--11186},
  year={1996},
  doi={10.1103/PhysRevB.54.11169},
  url={https://doi.org/10.1103/PhysRevB.54.11169}
}

@article{hutter2014cp2k,
  title={CP2K: atomistic simulations of condensed matter systems},
  author={Hutter, J{\"u}rg and Iannuzzi, Marcella and Schiffmann, Florian and VandeVondele, Joost},
  journal={WIREs Computational Molecular Science},
  volume={4},
  number={1},
  pages={15--25},
  year={2014},
  doi={10.1002/wcms.1159},
  url={https://doi.org/10.1002/wcms.1159}
}

@manual{cuda2019programming,
  title={CUDA C Programming Guide},
  author={{NVIDIA Corporation}},
  year={2019},
  note={\url{https://docs.nvidia.com/cuda/cuda-c-programming-guide/}}
}

@article{ufimtsev2009gpu2,
  title={Quantum Chemistry on Graphical Processing Units. 2. Direct Self-Consistent-Field Implementation},
  author={Ufimtsev, Ilya S. and Martinez, Todd J.},
  journal={Journal of Chemical Theory and Computation},
  volume={5},
  number={10},
  pages={2619--2628},
  year={2009},
  doi={10.1021/ct800526s},
  url={https://doi.org/10.1021/ct800526s}
}

@article{galvez2022eri_gpu,
  title={High-performance GPU-accelerated evaluation of electron repulsion integrals},
  author={G{\'a}lvez Vallejo, Jorge Luis and Barca, Giuseppe M. J. and Gordon, Mark S.},
  journal={Molecular Physics},
  year={2022},
  doi={10.1080/00268976.2022.2112987},
  url={https://doi.org/10.1080/00268976.2022.2112987}
}

@article{yasuda2008gpu,
  title={Accelerating Density Functional Calculations with Graphics Processing Unit},
  author={Yasuda, Koji},
  journal={Journal of Chemical Theory and Computation},
  volume={4},
  number={8},
  pages={1230--1236},
  year={2008},
  doi={10.1021/ct8001046},
  url={https://doi.org/10.1021/ct8001046}
}

@article{laqua2020snlink,
  title={Highly Efficient, Linear-Scaling Seminumerical Exact-Exchange Method for Graphic Processing Units},
  author={Laqua, Henryk and Thompson, Travis H. and Kussmann, J{\"o}rg and Ochsenfeld, Christian},
  journal={Journal of Chemical Theory and Computation},
  volume={16},
  number={3},
  pages={1456--1468},
  year={2020},
  doi={10.1021/acs.jctc.9b00860},
  url={https://doi.org/10.1021/acs.jctc.9b00860}
}

@article{kussmann2021highly,
  title={Highly Efficient Resolution-of-Identity Density Functional Theory Calculations on Central and Graphics Processing Units},
  author={Kussmann, J{\"o}rg and Laqua, Henryk and Ochsenfeld, Christian},
  journal={Journal of Chemical Theory and Computation},
  volume={17},
  number={3},
  pages={1512--1521},
  year={2021},
  doi={10.1021/acs.jctc.0c01252},
  url={https://doi.org/10.1021/acs.jctc.0c01252}
}

@article{vahtras1993ri,
  title={Integral approximations for LCAO-SCF calculations},
  author={Vahtras, Olav and Alml{\"o}f, Jan and Feyereisen, Mark W.},
  journal={Chemical Physics Letters},
  volume={213},
  number={5--6},
  pages={514--518},
  year={1993},
  doi={10.1016/0009-2614(93)89151-7},
  url={https://doi.org/10.1016/0009-2614(93)89151-7}
}

@article{williams2021achieving,
  title={Achieving performance portability in Gaussian basis set density functional theory on accelerator based architectures in NWChemEx},
  author={Williams-Young, David B. and Bagusetty, Abhishek and de Jong, Wibe A. and Doerfler, Douglas and van Dam, Hubertus J. J. and V{\'a}zquez-Mayagoitia, {\'A}lvaro and Windus, Theresa L. and Yang, Chao},
  journal={Parallel Computing},
  volume={108},
  pages={102829},
  year={2021},
  doi={10.1016/j.parco.2021.102829},
  url={https://doi.org/10.1016/j.parco.2021.102829}
}

@article{manathunga2020parallel,
  title={Parallel implementation of density functional theory methods in the quantum interaction computational kernel program},
  author={Manathunga, Madushanka and Miao, Yipu and Mu, Dawei and G{"o}tz, Andreas W. and Merz, Jr., Kenneth M.},
  journal={Journal of Chemical Theory and Computation},
  volume={16},
  number={7},
  pages={4315--4326},
  year={2020},
  doi={10.1021/acs.jctc.0c00290},
  url={https://doi.org/10.1021/acs.jctc.0c00290}
}

@article{ju2024acceleration,
  title={Acceleration without Disruption: DFT Software as a Service},
  author={Ju, Fusong and Wei, Xinran and Huang, Lin and Jenkins, Andrew J. and Xia, Leo and Zhang, Jia and Zhu, Jianwei and Yang, Han and Shao, Bin and Dai, Peggy and Williams-Young, David B. and Mayya, Ashwin and Hooshmand, Zahra and Efimovskaya, Alexandra and Baker, Nathan A. and Troyer, Matthias and Liu, Hongbin},
  journal={Journal of Chemical Theory and Computation},
  year={2024},
  doi={10.1021/acs.jctc.4c00940},
  url={https://doi.org/10.1021/acs.jctc.4c00940}
}

@article{anzt2022ginkgo,
  title={Ginkgo: A modern linear operator algebra framework for high performance computing},
  author={Anzt, Hartwig and Cojean, Terry and Flegar, Goran and G{\"o}bel, Fritz and Gr{\"u}tzmacher, Thomas and Nayak, Pratik and Ribizel, Tobias and Tsai, Yuhsiang Mike and Quintana-Ort{\'\i}, Enrique S.},
  journal={ACM Transactions on Mathematical Software},
  volume={48},
  number={1},
  pages={1--33},
  year={2022},
  doi={10.1145/3480935},
  url={https://doi.org/10.1145/3480935}
}

@article{becke1988multicenter,
  title={A multicenter numerical integration scheme for polyatomic molecules},
  author={Becke, Axel D.},
  journal={The Journal of Chemical Physics},
  volume={88},
  number={4},
  pages={2547--2553},
  year={1988},
  doi={10.1063/1.454033},
  url={https://doi.org/10.1063/1.454033}
}

@article{treutler1995grid,
  title={Efficient molecular numerical integration schemes},
  author={Treutler, Oliver and Ahlrichs, Reinhart},
  journal={The Journal of Chemical Physics},
  volume={102},
  number={1},
  pages={346--354},
  year={1995},
  doi={10.1063/1.469408},
  url={https://doi.org/10.1063/1.469408}
}

@article{gill1993grid,
  title={A standard grid for density functional calculations},
  author={Gill, Peter M. W. and Johnson, Brian G. and Pople, John A.},
  journal={Chemical Physics Letters},
  volume={209},
  number={5--6},
  pages={506--512},
  year={1993},
  doi={10.1016/0009-2614(93)80125-9},
  url={https://doi.org/10.1016/0009-2614(93)80125-9}
}

@article{lebedev1976grid,
  title={Quadratures on a sphere},
  author={Lebedev, V. I.},
  journal={USSR Computational Mathematics and Mathematical Physics},
  volume={16},
  number={2},
  pages={10--24},
  year={1976},
  doi={10.1016/0041-5553(76)90100-2},
  url={https://doi.org/10.1016/0041-5553(76)90100-2}
}

@inproceedings{vuduc2005fast,
  title={Fast sparse matrix-vector multiplication by exploiting variable block structure},
  author={Vuduc, Richard W and Moon, Hyun-Jin},
  booktitle={High Performance Computing and Communications: First International Conference, HPCC 2005, Sorrento, Italy, September 21-23, 2005. Proceedings 1},
  pages={807--816},
  year={2005},
  organization={Springer},
  doi={10.1007/11557654_91},
  url={https://doi.org/10.1007/11557654_91}
}

@inproceedings{eberhardt2016optimization,
  title={Optimization of block sparse matrix-vector multiplication on shared-memory parallel architectures},
  author={Eberhardt, Ryan and Hoemmen, Mark},
  booktitle={2016 IEEE International Parallel and Distributed Processing Symposium Workshops (IPDPSW)},
  pages={663--672},
  year={2016},
  organization={IEEE},
  doi={10.1109/IPDPSW.2016.42},
  url={https://doi.org/10.1109/IPDPSW.2016.42}
}

@misc{cublas,
    title={cuBLAS: The NVIDIA CUDA Basic Linear Algebra Subroutine Library},
    howpublished={\url{https://docs.nvidia.com/cuda/cublas/}},
    publisher={NVIDIA Corporation},
    year={2024},
    note={Accessed: 2024-11-11}
}

@inproceedings{cusparse,
  title={Cusparse library},
  author={Naumov, Maxim and Chien, L and Vandermersch, Philippe and Kapasi, Ujval},
  booktitle={GPU Technology Conference},
  volume={12},
  year={2010}
}

@inproceedings{qin2021extending,
  title={Extending sparse tensor accelerators to support multiple compression formats},
  author={Qin, Eric and Jeong, Geonhwa and Won, William and Kao, Sheng-Chun and Kwon, Hyoukjun and Srinivasan, Sudarshan and Das, Dipankar and Moon, Gordon E and Rajamanickam, Sivasankaran and Krishna, Tushar},
  booktitle={2021 IEEE International Parallel and Distributed Processing Symposium (IPDPS)},
  pages={1014--1024},
  year={2021},
  organization={IEEE},
  doi={10.1109/IPDPS49936.2021.00110},
  url={https://doi.org/10.1109/IPDPS49936.2021.00110}
}

@inproceedings{steinberger2014softshell,
  title={Whippletree: task-based scheduling of dynamic workloads on the {GPU}},
  author={Steinberger, Markus and Kenzel, Michael and Boechat, Pedro and Kerber, Bernhard and Dokter, Mark and Schmalstieg, Dieter},
  booktitle={ACM Transactions on Graphics (TOG)},
  volume={33},
  number={6},
  pages={1--11},
  year={2014},
  doi={10.1145/2661229.2661250},
  url={https://doi.org/10.1145/2661229.2661250}
}

@article{bulucc2012parallel,
  title={Parallel sparse matrix-matrix multiplication and indexing: Implementation and experiments},
  author={Bulu{\c{c}}, Aydin and Gilbert, John R},
  journal={SIAM Journal on Scientific Computing},
  volume={34},
  number={4},
  pages={C170--C191},
  year={2012},
  publisher={SIAM},
  doi={10.1137/110848244},
  url={https://doi.org/10.1137/110848244}
}

@article{artemov2021sparse,
  title={Sparse approximate matrix-matrix multiplication for density matrix purification with error control},
  author={Artemov, Anton G and Rubensson, Emanuel H},
  journal={Journal of Computational Physics},
  volume={438},
  pages={110354},
  year={2021},
  publisher={Elsevier},
  doi={10.1016/j.jcp.2021.110354},
  url={https://doi.org/10.1016/j.jcp.2021.110354}
}

@article{zhou2020gpu,
  title={A GPU implementation of classical density functional theory for rapid prediction of gas adsorption in nanoporous materials},
  author={Zhou, Musen and Wu, Jianzhong},
  journal={The Journal of Chemical Physics},
  volume={153},
  number={7},
  year={2020},
  publisher={AIP Publishing},
  doi={10.1063/5.0020797},
  url={https://doi.org/10.1063/5.0020797}
}

@article{barca2020recent,
  title={Recent developments in the general atomic and molecular electronic structure system},
  author={Barca, Giuseppe M. J. and Bertoni, Colleen and Carrington, Laura and Datta, Dipayan and De Silva, Nuwan and Deustua, J. Emiliano and Fedorov, Dmitri G. and Gour, Jeffrey R. and Gunber, Anastasia O. and Guidez, Emilie and others},
  journal={The Journal of Chemical Physics},
  volume={152},
  number={15},
  pages={154102},
  year={2020},
  doi={10.1063/5.0005188},
  url={https://doi.org/10.1063/5.0005188}
}

@article{gotz2012routine,
  title={Routine microsecond molecular dynamics simulations with {AMBER} on {GPUs}. 1. Generalized Born},
  author={G{\"o}tz, Andreas W. and Williamson, Mark J. and Xu, Dong and Poole, Duncan and Le Grand, Scott and Walker, Ross C.},
  journal={Journal of Chemical Theory and Computation},
  volume={8},
  number={5},
  pages={1542--1555},
  year={2012},
  doi={10.1021/ct200909j},
  url={https://doi.org/10.1021/ct200909j}
}

@article{haidar2018harnessing,
  title={Harnessing {GPU} tensor cores for fast {FP16} arithmetic to speed up mixed-precision iterative refinement solvers},
  author={Haidar, Azzam and Tomov, Stanimire and Dongarra, Jack and Higham, Nicholas J.},
  journal={Proceedings of the International Conference for High Performance Computing, Networking, Storage, and Analysis (SC '18)},
  pages={1--11},
  year={2018},
  doi={10.1109/SC.2018.00050},
  url={https://doi.org/10.1109/SC.2018.00050}
}

@article{weigend2005balanced,
  title={Balanced basis sets of split valence, triple zeta valence and quadruple zeta valence quality for H to Rn: Design and assessment of accuracy},
  author={Weigend, Florian and Ahlrichs, Reinhart},
  journal={Physical Chemistry Chemical Physics},
  volume={7},
  number={18},
  pages={3297--3305},
  year={2005},
  publisher={Royal Society of Chemistry},
  doi={10.1039/b508541a},
  url={https://doi.org/10.1039/b508541a}
}

@article{zhao2008m06,
  title={The M06 suite of density functionals for main group thermochemistry, thermochemical kinetics, noncovalent interactions, excited states, and transition elements: two new functionals and systematic testing of four M06-class functionals and 12 other functionals},
  author={Zhao, Yan and Truhlar, Donald G},
  journal={Theoretical chemistry accounts},
  volume={120},
  number={1},
  pages={215--241},
  year={2008},
  publisher={Springer},
  doi={10.1007/s00214-007-0310-x},
  url={https://doi.org/10.1007/s00214-007-0310-x}
}

@article{stocks2025efficient,
  title={Efficient algorithms for GPU accelerated evaluation of the DFT exchange-correlation functional},
  author={Stocks, Ryan and Barca, Giuseppe MJ},
  journal={Journal of Chemical Theory and Computation},
  volume={21},
  number={20},
  pages={10263--10280},
  year={2025},
  publisher={ACS Publications},
  doi={10.1021/acs.jctc.5c01229},
  url={https://doi.org/10.1021/acs.jctc.5c01229}
}

@article{wu2025enhancing,
  title={Enhancing GPU-Acceleration in the Python-Based Simulations of Chemistry Frameworks},
  author={Wu, Xiaojie and Sun, Qiming and Pu, Zhichen and Zheng, Tianze and Ma, Wenzhi and Yan, Wen and Xia, Yu and Wu, Zhengxiao and Huo, Mian and Li, Xiang and others},
  journal={Wiley Interdisciplinary Reviews: Computational Molecular Science},
  volume={15},
  number={2},
  pages={e70008},
  year={2025},
  publisher={Wiley Online Library},
  doi={10.1002/wcms.70008},
  url={https://doi.org/10.1002/wcms.70008}
}

@inproceedings{markidis2018nvidia,
  title={Nvidia tensor core programmability, performance \& precision},
  author={Markidis, Stefano and Der Chien, Steven Wei and Laure, Erwin and Peng, Ivy Bo and Vetter, Jeffrey S},
  booktitle={2018 IEEE international parallel and distributed processing symposium workshops (IPDPSW)},
  pages={522--531},
  year={2018},
  organization={IEEE},
  doi={10.1109/IPDPSW.2018.00091},
  url={https://doi.org/10.1109/IPDPSW.2018.00091}
}

@article{choquette2021nvidia,
  title={Nvidia a100 tensor core gpu: Performance and innovation},
  author={Choquette, Jack and Gandhi, Wishwesh and Giroux, Olivier and Stam, Nick and Krashinsky, Ronny},
  journal={IEEE Micro},
  volume={41},
  number={2},
  pages={29--35},
  year={2021},
  publisher={IEEE},
  doi={10.1109/MM.2021.3061394},
  url={https://doi.org/10.1109/MM.2021.3061394}
}

@inproceedings{volkov2008benchmarking,
  title={Benchmarking {GPUs} to tune dense linear algebra},
  author={Volkov, Vasily and Demmel, James W.},
  booktitle={SC '08: Proceedings of the 2008 ACM/IEEE Conference on Supercomputing},
  pages={1--11},
  year={2008},
  organization={IEEE},
  doi={10.1109/SC.2008.5214359},
  url={https://doi.org/10.1109/SC.2008.5214359}
}

@article{nickolls2010gpu,
  title={{GPU} computing},
  author={Nickolls, John and Dally, William J.},
  journal={Proceedings of the IEEE},
  volume={98},
  number={8},
  pages={1479--1492},
  year={2010}
}

@inproceedings{bell2009implementing,
  title={Implementing sparse matrix-vector multiplication on throughput-oriented processors},
  author={Bell, Nathan and Garland, Michael},
  booktitle={Proceedings of the Conference on High Performance Computing Networking, Storage and Analysis (SC '09)},
  pages={1--11},
  year={2009},
  doi={10.1145/1654059.1654078},
  url={https://doi.org/10.1145/1654059.1654078}
}

@article{filippone2017sparse,
  title={Sparse matrix-vector multiplication on {GPGPUs}},
  author={Filippone, Salvatore and Cardellini, Valeria and Barbieri, Davide and Luque, Alex},
  journal={ACM Transactions on Mathematical Software},
  volume={43},
  number={4},
  pages={1--49},
  year={2017},
  doi={10.1145/3017994},
  url={https://doi.org/10.1145/3017994}
}

@inproceedings{yang2018design,
  title={Design principles for sparse matrix multiplication on the {GPU}},
  author={Yang, Carl and Bulu{\c{c}}, Aydin and Owens, John D.},
  booktitle={Euro-Par 2018: Parallel Processing},
  pages={672--687},
  year={2018},
  publisher={Springer},
  doi={10.1007/978-3-319-96983-1_48},
  url={https://doi.org/10.1007/978-3-319-96983-1_48}
}

@article{davis2011university,
  title={The {University of Florida} sparse matrix collection},
  author={Davis, Timothy A. and Hu, Yifan},
  journal={ACM Transactions on Mathematical Software},
  volume={38},
  number={1},
  pages={1--25},
  year={2011},
  doi={10.1145/2049662.2049663},
  url={https://doi.org/10.1145/2049662.2049663}
}

\newpage

\appendices

\section{Supplementary Technical Details}
\label{sec:appendix_technical}

\subsection{DFT Primer for the HPC Audience}
\label{sec:dft_primer}

This appendix provides a concise introduction to the quantum-chemistry concepts
referenced in the main text, aimed at readers familiar with GPU computing
but not with electronic-structure theory.

\paragraph{The Kohn--Sham framework.}
Density Functional Theory (DFT) computes the ground-state electronic
properties of a molecular or solid-state system by solving for the electron
density $\rho(\mathbf{r})$ rather than the full many-body wavefunction.
The Kohn--Sham (KS) formulation~\cite{kohnsham1965self} replaces the
intractable many-electron Schr\"odinger equation with a set of
single-particle equations. The solutions of which, called the KS orbitals
$\psi_k(\mathbf{r})$, reproduce the exact ground-state density. The
orbitals are expanded in a finite set of \emph{basis functions}
$\{\chi_\mu\}$, giving the matrix eigenvalue problem
$\mathbf{F}\mathbf{C}=\mathbf{S}\mathbf{C}\boldsymbol{\epsilon}$ stated
in Section~II. The Fock matrix $\mathbf{F}$ depends on the orbitals
through the density, so the equation is solved iteratively in a
\emph{self-consistent field} (SCF) loop until the input and output
densities converge.

\paragraph{Gaussian basis functions.}
In the molecular DFT codes, each basis function
$\chi_\mu(\mathbf{r})$ is a \emph{contracted Gaussian}: a fixed linear
combination of Gaussian primitives
$g(\mathbf{r}) = x^l y^m z^n \exp(-\alpha|\mathbf{r}-\mathbf{R}|^2)$
centered on atom $\mathbf{R}$.
Two properties are computationally important:
(i)~products of two Gaussians centered on different atoms are
themselves Gaussians, making integral evaluation tractable; and
(ii)~each Gaussian decays exponentially with distance from its center,
so its value on a grid point far from~$\mathbf{R}$ is negligibly small.
Property~(ii) is the physical origin of the locality-driven sparsity
exploited by KerneLDI.

\paragraph{The density matrix.}
The density matrix $\mathbf{D}$ is constructed from the occupied KS
orbitals: $D_{\mu\nu}=\sum_{k\in\mathrm{occ}} C_{\mu k}C_{\nu k}$,
where $C_{\mu k}$ is the expansion coefficient of orbital~$k$ in basis
function~$\mu$. It fully determines the electron density on any grid
point via $\rho(\mathbf{r}_i)=\sum_{\mu\nu}D_{\mu\nu}\chi_\mu(\mathbf{r}_i)\chi_\nu(\mathbf{r}_i)$.

\paragraph{Numerical integration grids.}
The exchange--correlation energy $E_{\text{xc}}[\rho]$ is a functional of
the density whose analytic integral is generally unknown, so it must be
evaluated by numerical quadrature over a three-dimensional grid.
Following Becke~\cite{becke1988multicenter}, molecular grids are
constructed as a superposition of atom-centered grids, each consisting
of a radial component (e.g., Gauss--Chebyshev or Euler--Maclaurin) and
an angular component (Lebedev spherical quadrature~\cite{lebedev1976grid}).
A Becke partition function assigns each grid point to an atom so that
the molecular integral decomposes into a sum of atomic contributions.
The ``Level-4 grid'' referenced in this paper corresponds to PySCF's
grid convention with up to $\sim$590 angular points per radial shell per atom,
yielding on the order of one to several million grid points for a
100-atom system depending on the pruning scheme.

\paragraph{Grid pruning.}
Because Gaussian basis functions decay rapidly with distance, most basis
functions have negligible values at grid points far from their center.
\emph{Grid pruning} discards basis-function/grid-point pairs
$(\mu,\,i)$ for which $|\chi_\mu(\mathbf{r}_i)| < \epsilon_{\text{grid}}$,
typically with $\epsilon_{\text{grid}} \approx 10^{-10}$ to $10^{-14}$.
This element-wise pruning determines which entries of the
basis-on-grid matrices $\boldsymbol{\Phi}$ and $\boldsymbol{\Psi}$ are
nonzero. KerneLDI operates at a coarser granularity: after the pruned
matrices are partitioned into $d\times d$ blocks, the block-level
filtering thresholds $t_b$ and $t_d$ further discard blocks and block
pairs whose contributions are negligible (see Sections~III-B and~IV).

\paragraph{The constant $b$ in Eq.~\eqref{eq:fxc_matrix_form}.}
The factor $b$ in the EXC matrix formula accounts for the spin
multiplicity: $b=1$ for restricted (closed-shell) calculations and $b=2$
for unrestricted (open-shell) calculations in which $\alpha$ and $\beta$
spin contributions are handled separately.

\subsection{Reordering Implementation Details}
\label{sec:reordering_details}

This subsection provides the implementation details of the locality-preserving reordering described in Section~III-B.

\subsubsection{Grid-Point Morton Ordering}

Grid points are reordered using a Z-order (Morton) space-filling curve to concentrate spatially nearby points into contiguous memory ranges. Each grid point with Cartesian coordinates $(x, y, z)$ is first scaled to an integer lattice by multiplying by a resolution factor (128 in our implementation) and truncating to integer values. A 60-bit Morton code is then constructed by bit-interleaving the three 20-bit integer coordinates:
\begin{equation}
  m = \sum_{k=0}^{19} \bigl( z_k \cdot 2^{3k} + y_k \cdot 2^{3k+1} + x_k \cdot 2^{3k+2} \bigr),
  \label{eq:morton_code}
\end{equation}
where $x_k$, $y_k$, $z_k$ denote the $k$-th bits of the discretized coordinates. Grid points are then sorted by their Morton codes using a stable sort, which preserves the relative order of points sharing the same code. This mapping places three-dimensionally adjacent grid points at nearby one-dimensional positions, reducing fragmentation when the basis-on-grid matrices are subsequently partitioned into fixed-size blocks.

\subsubsection{Overlap-Signature Construction and Basis-Function Clustering}

Basis functions are reordered by clustering their overlap signatures,
as outlined in Section~III-B. The overlap matrix
$\mathbf{S}$ (Eq.~\eqref{eq:overlap_matrix}) is already available from
the Kohn--Sham setup, so no additional integral evaluation is required.
For each basis function~$i$, the overlap signature
$\mathbf{s}_i = (S_{i1}, S_{i2}, \dots, S_{iN_{\mathrm{basis}}})$
(Eq.~\eqref{eq:overlap_signature}) is extracted as the $i$-th row
of~$\mathbf{S}$.

A pairwise cosine-distance matrix is then formed:
\begin{equation}
  d(i,j) = 1 - \frac{\mathbf{s}_i \cdot \mathbf{s}_j}
                     {\|\mathbf{s}_i\|\,\|\mathbf{s}_j\|},
  \label{eq:cosine_distance}
\end{equation}
which equals zero for identical signatures and approaches one for
orthogonal ones.

Agglomerative (hierarchical) clustering with average linkage is applied
to the distance matrix. Starting from $N_{\mathrm{basis}}$ singleton
clusters, the algorithm repeatedly merges the pair with the smallest
average inter-cluster distance until a single cluster remains, producing
a binary dendrogram. At each merge the two sub-trees are oriented so
that their mutually closest leaves are adjacent, yielding a leaf ordering
that places basis functions with similar overlap signatures---and hence
similar spatial support---at neighbouring index positions.

The resulting leaf order is read off as the basis-function permutation
$\pi$, and the operand matrices are reindexed accordingly. Because the
overlap matrix is symmetric and typically has $O(N_{\mathrm{basis}})$
significant entries per row (due to Gaussian decay), the clustering runs
in $O(N_{\mathrm{basis}}^2 \log N_{\mathrm{basis}})$ time and is
executed once per molecular geometry as part of the preprocessing stage
described in Section~III-D.

\subsection{BCS Data Layout Specification}
\label{sec:bcs_layout}

The Block Compressed Sparse (BCS) layouts used by KerneLDI are
block-granularity analogues of the well-known Compressed Sparse Row (CSR)/ Compressed Sparse Column (CSC) element-wise sparse
formats:

\paragraph{BCS(R) --- Block Compressed Sparse Row.}
The retained blocks of matrix $\mathbf{A}$ are grouped by block-row.  The
format consists of three arrays:
\begin{itemize}
\item \texttt{row\_ptr}$[0 \ldots M_r]$: the $i$-th entry gives the
      offset into \texttt{col\_idx} (and the data array) where
      block-row~$i$ begins.
\item \texttt{col\_idx}$[0 \ldots R_A{-}1]$: for each retained block,
      its block-column index.
\item \texttt{data}$[0 \ldots R_A \cdot d^2{-}1]$: the dense $d\times d$
      entries of each retained block, stored contiguously in row-major order
      within each block.
\end{itemize}
Blocks sharing the same block-row are stored adjacent in memory, so
iterating over all blocks in a given row requires a single contiguous
read of length
$(\texttt{row\_ptr}[i{+}1]-\texttt{row\_ptr}[i])\times d^2$ elements.

\paragraph{BCS(C) --- Block Compressed Sparse Column.}
The retained blocks of matrix $\mathbf{B}$ are grouped by block-column
using \texttt{col\_ptr}, \texttt{row\_idx}, and \texttt{data} arrays with
mirrored semantics. This ensures that the multiplier can access all blocks
contributing to a single output block-column with coalesced reads.

\paragraph{Comparison with element-wise formats.}
In standard CSR, every nonzero carries one column index (typically 4
bytes), yielding metadata overhead $\approx 4\times\mathit{nnz}$ bytes.
In BCS, each retained \emph{block} carries one column index, so the
metadata overhead is $\approx 4\times R$ bytes for $R$ retained blocks,
while the data volume is $R \times d^2 \times 4$ bytes (single precision).
For the typical block size $d=32$, the metadata-to-data ratio is
$1\!:\!d^2 = 1\!:\!1024$, which is negligible compared to element-wise
formats.

\subsection{Parallel Addressing}
\label{sec:para_add}

The parallel addressing scheme described here is invoked in the dense block
multiplier (Section~III-C) immediately after
the block-pair filtering step (Eq.~\eqref{eq:block_pair_filter}). At that point, each
CUDA thread has independently evaluated whether its assigned block pair
passes the significance threshold, producing a per-thread Boolean result.
The multiplier must then \emph{compactly enumerate} only the retained pairs
so that subsequent shared-memory loading and Tensor Core computation operate
on a contiguous, gap-free work list. A na\"ive serial scan would serialize
this step; KerneLDI instead uses a parallel prefix sum to perform the
compaction entirely within shared memory in $O(\log n)$ steps, which is why
we detail the procedure here.

In the algorithm, the input is a 0/1 array $\mathbf{M}$ of length equal to the total
number of threads, stored in shared memory, with $\mathbf{M}[\mathit{tid}]=1$
if thread $\mathit{tid}$'s block pair passed the filter. A parallel
\emph{exclusive prefix sum} over $\mathbf{M}$ yields an array
$\mathbf{P}$ of length $\mathit{thread\_num}+1$ whose $i$-th entry
equals the number of retained pairs in positions $0$ through~$i-1$.
For any position where $\mathbf{M}[i]=1$,
the corresponding prefix sum $\mathbf{P}[i]$ directly gives that pair's
compact output index, allowing all retained pairs to be written to a
dense output array $\mathbf{I}$ without gaps or atomic operations.

\begin{algorithm}
\caption{Parallel Addressing}
\label{alg:parallel_where}
\begin{algorithmic}[1]
\State \textbf{Parameters:} Number of threads $thread\_num$
\State \textbf{Input:} Mask array after filtering $\mathbf{M} \in [0,1]^{thread\_num}$
\State \textbf{Output:} Total count $S$, index array $\mathbf{I} \in \mathbb{N}^{S}$
\State Initialize array $\mathbf{P} \in \mathbb{N}^{thread\_num+1}$ in shared memory
\State Copy $\mathbf{M}$ to $\mathbf{P}$ in each thread \Comment{Thread-level assignment}
\State Synchronize threads
\State Update $\mathbf{P}$ using prefix sum algorithm \Comment{Thread-level computation}
\State Synchronize threads
\State $S \gets \mathbf{P}[thread\_num]$
\State $v \gets \mathbf{P}[thread\_id + 1]$
\If{$\mathbf{P}[thread\_id] \neq v$}
    \State $\mathbf{I}[v - 1] \gets thread\_id$
\EndIf
\end{algorithmic}
\end{algorithm}

\subsection{Tensor Core Hardware Mapping}
\label{sec:tc_mapping}

KerneLDI invokes Tensor Cores via the WMMA API~\cite{markidis2018nvidia}
with FP16 input fragments and FP32 accumulation. Each warp loads a
$16\times 16$ sub-tile from shared memory into WMMA fragments, executes
a hardware matrix multiply-accumulate, and writes the FP32 result back
to a shared-memory accumulator. For a $32\times 32$ block pair, four
such tiles are computed and accumulated. Because Tensor Cores deliver
$\sim$125\,TFLOPS (FP16) on V100 versus $\sim$15.7\,TFLOPS for FP32
CUDA cores, this mapping yields a significant throughput gain on the
dense sub-problems, while the FP32 accumulation preserves the numerical
accuracy demonstrated in Section~IV.

\section{Extended Related Work}
\label{sec:extended_related_work}

This supplement provides a more detailed discussion of related work
summarized in the main text.

\paragraph{Structured and block-sparse matrix computation on GPUs.}
Generic sparse libraries such as cuSPARSE~\cite{cusparse} provide
element-wise formats (CSR, COO) optimized for highly sparse regimes, and
autotuning frameworks have further improved SpMV and SpMM
throughput~\cite{vuduc2005fast,eberhardt2016optimization}. For workloads
with coarser structure, block-sparse representations have been explored
in accelerator-oriented sparse tensor
processing~\cite{qin2021extending} and in quantum-chemistry
density-matrix methods~\cite{artemov2021sparse}. NVIDIA's Ampere
architecture introduced hardware-level 2:4 structured
sparsity~\cite{choquette2021nvidia}, which doubles effective throughput
for qualifying weight matrices but requires a fixed fine-grained
pattern. These approaches target either very high sparsity or
predetermined structures. KerneLDI addresses the intermediate regime
where block occupancy varies dynamically across grid batches, and the
sparsity pattern is determined at runtime by spatial locality and
screening thresholds.

\paragraph{GPU-accelerated numerical integration in DFT.}
GPU acceleration of Kohn--Sham DFT has progressed from early GPU-based
electron repulsion integral
evaluation~\cite{ufimtsev2009gpu2,yasuda2008gpu} to density-fitting
approaches for Coulomb and exchange
terms~\cite{galvez2022eri_gpu}. Parallel GPU implementations have also
been demonstrated for classical density functional
theory~\cite{zhou2020gpu}. For EXC integration specifically,
Williams-Young et al.\ demonstrated efficient GPU execution within the
GauXC framework by grouping grid batches into dense sub-matrices and
dispatching them via batched
GEMM~\cite{williams2021achieving}, and Manathunga et al.\ adopted a
similar batched dense strategy in the QUICK
package~\cite{manathunga2020parallel}. More recent work by
Wu et al.~\cite{wu2025enhancing} and
Stocks et al.~\cite{stocks2025efficient} has further improved
batched-GEMM-based EXC pipelines through refined grid partitioning and
memory management. All of these approaches regularize EXC into
uniformly shaped dense workloads, discarding the block-level sparsity
that KerneLDI preserves.

\paragraph{Tensor Core utilization in scientific computing.}
NVIDIA Tensor Cores, originally designed for deep-learning
workloads~\cite{markidis2018nvidia}, have been adopted in scientific
computing for mixed-precision dense linear algebra, iterative solvers,
and FFT-based computations~\cite{haidar2018harnessing}.
Barca et al.~\cite{barca2020recent} demonstrated the use of Tensor
Cores for accelerating integral evaluation in quantum chemistry.
KerneLDI leverages Tensor Cores within its dense block multiplier to
accelerate retained block-pair products, adapting hardware-accelerated
warp-level matrix multiply-accumulate to the block-filtered execution
model.

\paragraph{Dynamic task scheduling for irregular GPU workloads.}
Load imbalance arising from irregular or data-dependent work
distributions is a well-known challenge in GPU computing. Task-based
runtime systems and work-stealing schedulers have been proposed for GPU
workload management~\cite{steinberger2014softshell}, molecular dynamics
force decomposition~\cite{gotz2012routine}, and sparse linear
algebra~\cite{anzt2022ginkgo}. Within DFT, grid-based numerical
integration is inherently heterogeneous because different spatial
regions retain different numbers of significant basis functions.
KerneLDI addresses this heterogeneity through a pull-based dynamic
scheduling strategy that assigns filtered grid batches to GPU devices
on demand, ensuring balanced utilization across multi-GPU platforms
without requiring a priori cost estimates.

\end{document}


\twocolumn[%
{\begin{center}
\Huge
Appendix: Artifact Description/Artifact Evaluation
\end{center}}
]


\appendixAD

\section{Overview of Contributions and Artifacts}

\subsection{Paper's Main Contributions}

\begin{description}
\item[$C_1$] KerneLDI framework: a block-filtered representation and GPU dense block multiplier for locality-driven integration in quantum chemistry, with Tensor Core acceleration.
\item[$C_2$] A dynamic multi-GPU execution model with pull-based task scheduling for load-balanced EXC evaluation across heterogeneous grid batches.
\item[$C_3$] Empirical evaluation on exchange--correlation (EXC) integration: up to $10\times$ single-GPU speedup, $1.5\times$--$2.7\times$ end-to-end SCF acceleration over GPU4PySCF, and $\sim$$6\times$ AIMD throughput improvement.
\end{description}

\subsection{Computational Artifacts}

\begin{description}
\item[$A_1$] Source code implementing the KerneLDI framework, including the block-filtered formatter, GPU dense block multiplier, multi-GPU scheduler, and the full SCF driver. Includes a Docker environment for reproducible builds.
The repository is currently private and will be made publicly available upon paper acceptance. A Zenodo DOI will be assigned at that time.
\url{https://github.com/bjzgcai/megadft}

\item[$A_2$] Molecular structure files (\texttt{.xyz} format) and basis set definitions used in all experiments. The six representative molecules and additional test systems are bundled in the \texttt{sample/} directory of $A_1$. The 329-molecule benchmark dataset is from Ju et~al.~(JCTC, 2024; DOI: \url{https://doi.org/10.1021/acs.jctc.4c00940}).
\end{description}

\begin{center}
\begin{tabular}{rll}
\toprule
Artifact  &  Contributions  &  Related Paper \\
          &  Supported      &  Elements \\
\midrule
$A_1$  &  $C_1$, $C_2$, $C_3$  &  Figures~4--8 \\
\midrule
$A_2$  &  $C_3$                 &  Figures~4--8 \\
       &                        &  (input data) \\
\bottomrule
\end{tabular}
\end{center}

\section{Artifact Identification}

\newartifact  

\artrel

$A_1$ contains the complete source code for all algorithmic contributions ($C_1$, $C_2$) and produces every experimental result in the paper ($C_3$). $A_2$ provides the molecular inputs required by $A_1$.

\artexp

Running $A_1$ on the molecular systems in $A_2$ should reproduce:
\begin{itemize}
    \item EXC energy and gradient matching the dense cuBLAS baseline within MAE $\sim$\!$10^{-11}$--$10^{-9}$~Hartree (Fig.~4).
    \item Single-GPU EXC speedup growing with system size, up to $\sim$\!$10\times$ on the largest molecules (Figs.~4,~5).
    \item End-to-end SCF wall-clock times $1.5\times$--$2.7\times$ faster than GPU4PySCF across six test molecules (Figure~6).
    \item Multi-GPU parallel efficiency of $\sim$79\% at 64~GPUs (Figure~7).
    \item AIMD trajectory reaching $\sim$35~ps vs.\ $\sim$6~ps (GPU baseline) within the same 800-hour wall-clock budget (Figure~8).
\end{itemize}

\arttime

Individual experiment runtimes (on one V100 GPU unless noted):
\begin{itemize}
    \item \textbf{329-molecule benchmark} (Section~IV-A, Figure~4): Each molecule requires one EXC energy + gradient evaluation per method. Total for KerneLDI across 329 molecules: $<$5~min. Including cuBLAS baseline: $<$30~min.
    \item \textbf{EXC single evaluation} (Section~IV-B, Figure~5): 3--70~ms per molecule with KerneLDI (energy); 8--70~ms (gradient). The largest system (crambin, 642 atoms) takes $\sim$70~ms. Running all six molecules $\times$ 3 methods $\times$ 3 repeats: $<$1~min.
    \item \textbf{End-to-end SCF} (Section~IV-C, Figure~6): Per-molecule SCF times range from $\sim$50~s (porphy) to $\sim$3000~s (crambin) with KerneLDI. Running all six molecules with both methods: $\sim$5~hours total.
    \item \textbf{Multi-GPU scaling} (Section~IV-D, Figure~7): Ubiquitin on 8/16/32/64 GPUs $\times$ 2 configurations (with/without dynamic scheduling) $\times$ 3 repeats. Estimated: $\sim$2~hours with 64 GPUs available.
    \item \textbf{AIMD simulation} (Section~IV-E, Figure~8): $\sim$800~GPU-hours for the full 35~ps trajectory with KerneLDI. This experiment is not reproducible within the 8-hour AE budget; a short validation run ($\sim$100 MD steps) can verify functional correctness in $<$1~hour.
\end{itemize}

\textbf{Summary:} Figures~4--7 can be reproduced within $\sim$6~hours on a single V100 GPU (excluding multi-GPU experiments). Figure~8 requires $\sim$800~GPU-hours for the full result; a reduced validation is feasible within 1~hour.

\artin

\artinpart{Hardware}

\begin{itemize}
    \item \textbf{GPU (used for KernelLDI and other GPU baselines):} NVIDIA Tesla V100-PCIE-16GB (5120 CUDA cores, 640 Tensor Cores, 16~GB HBM2).
    \item \textbf{CPU (baseline only):} AMD EPYC 7452 (32 cores, 2.35~GHz) with OpenBLAS.
    \item \textbf{Multi-GPU (Figure~7):} 8--64 V100 GPUs across multiple nodes, NVLink intra-node, InfiniBand inter-node.
    \item \textbf{Alternative GPUs:} The artifact has also been verified on NVIDIA A100 and H100 GPUs. Results are qualitatively consistent with V100; absolute speedup ratios may differ slightly due to different Tensor Core throughput and memory bandwidth, but the overall trends and conclusions hold.
\end{itemize}

\artinpart{Software}

A Docker image is provided (see Installation below) that bundles all dependencies. The key components and their versions are:

\begin{itemize}
    \item NVIDIA CUDA 12.6 (base image: \texttt{nvidia/cuda:}\linebreak\texttt{12.6.3-devel-ubuntu24.04})
    \item Libint 2.7.2 --- electron repulsion integrals
    \item Libxc 5.2.3 --- exchange--correlation functionals (with CUDA patch)
     \item OpenBLAS 0.3.26 --- optimised BLAS/LAPACK routines
    \item OpenMPI 4.1.6 --- MPI runtime
    \item UCX 1.18.0 + UCC 1.3.0 --- high-performance communication
    \item NVSHMEM 3.2.5 --- GPU-side shared memory
    \item cuSOLVERMp 0.6.0 + cuBLASMp 0.4.0 --- distributed GPU linear algebra
    \item Boost, Eigen 3, fmt, spdlog, CLI11, nlohmann-json --- C++ utilities
    \item CMake $\geq$ 3.21, GCC, \texttt{nvcc} (C++17)
\end{itemize}

For the end-to-end SCF baseline comparison (Figure~6):
\begin{itemize}
    \item GPU4PySCF (\url{https://github.com/pyscf/gpu4pyscf})
\end{itemize}

\artinpart{Datasets / Inputs}

\begin{itemize}
    \item \textbf{Molecular structures:} The six representative molecules (porphy, taxol, chignolin, trapcage, olestra, crambin), ubiquitin, and AceAla15Lys are provided as \texttt{.xyz} files in the \texttt{sample/} directory of $A_1$.
    \item \textbf{329-molecule dataset} (Figure~4): From Ju et~al.\ (JCTC, 2024; DOI: \url{https://doi.org/10.1021/acs.jctc.4c00940}). Structures contain $\geq$100 atoms and $\geq$4 distinct element types each.
    \item \textbf{Basis sets:} def2-SVP and def2-TZVP are pre-packaged in \texttt{resources/\allowbreak basis-set/} of $A_1$ (JSON format). Extra basis sets can be fetched via \texttt{scripts/\allowbreak download\_basis.sh}.
\end{itemize}

\artinpart{Installation and Deployment}

The recommended installation uses Docker to ensure a reproducible environment:

\begin{enumerate}
    \item Clone the repository: \\
    \url{https://github.com/bjzgcai/megadft.git}
    \item Build the Docker development image: \\
    \texttt{bash docker/build\_dev.sh}
    \item Start the Docker container: \\
    \texttt{bash docker/dev\_start.sh}
    \item Inside the container, compile: \\
    \texttt{cmake -B build -DCMAKE\_BUILD\_TYPE=Release} \\
    \texttt{cmake --build build -j \$(nproc)}
    \item Verify with a quick test: \\
    \texttt{./build/src/ath --mol sample/h2o.xyz \textbackslash} \\
    \texttt{~~~~--basis-set 6-31g* --verbose 6}
\end{enumerate}

For multi-GPU experiments, multiple containers or a bare-metal installation with the same dependency stack is needed, with MPI configured across nodes.

\artcomp

Experiments map to paper figures as follows. All use the M06-2X functional. Timing results are averaged over 3 runs.

\begin{description}
    \item[$T_1$] \textbf{Build} ($\sim$30~min): Build Docker image and compile the artifact as described above.

    \item[$T_2$] \textbf{EXC Accuracy \& Speedup} (Figures~4, 5; $<$30~min): \\
    Run the EXC energy and gradient evaluation on the six representative molecules with def2-SVP, comparing KerneLDI against the cuBLAS and OpenBLAS baselines. \\
    Parameters: block size $d=32$, $t_d=10^{-12}$, PySCF level-4 grid. \\
    Command pattern: \\
    \texttt{./build/src/ath --mol sample/<mol>.xyz \textbackslash} \\
    \texttt{~~~~--basis-set def2-svp --xc m06-2x}

    \item[$T_3$] \textbf{329-Molecule Benchmark} (Figure~4; $<$30~min): \\
    Same as $T_2$ but on the full 329-molecule dataset.

    \item[$T_4$] \textbf{End-to-End SCF} (Figure~6; $\sim$5~hours): \\
    Run full SCF convergence on the six molecules with def2-TZVP using both the artifact and GPU4PySCF. SCF convergence threshold: $10^{-8}$~Hartree. \\
    The largest single run (crambin) takes $\sim$50~min with KerneLDI.

    \item[$T_5$] \textbf{Multi-GPU Scaling} (Figure~7; $\sim$2~hours): \\
    Run ubiquitin with 8, 16, 32, 64 GPUs using \texttt{mpirun}. Run twice per GPU count: with and without dynamic scheduling.

    \item[$T_6$] \textbf{AIMD} (Figure~8; $\sim$800~GPU-hours): \\
    Run ab initio MD for AceAla15Lys for 35~ps simulated time. \\
    \emph{Reduced validation:} Run 100 MD steps ($<$1~hour) to verify energy conservation and gradient correctness.
\end{description}

Dependencies: $T_1 \rightarrow \{T_2, T_3, T_4, T_5, T_6\}$. Tasks $T_2$--$T_6$ are independent of each other.

\artout

\begin{itemize}
    \item \textbf{$T_2$/$T_3$ output:} Per-molecule EXC energy (Hartree), gradient norm, and wall-clock time (ms) for each method. Correctness check: KerneLDI energy should agree with cuBLAS within MAE $<10^{-9}$~Hartree. KerneLDI times for the six scaling molecules should range from $\sim$3~ms (porphy) to $\sim$70~ms (crambin) for energy, and $\sim$8--70~ms for gradient.
    \item \textbf{$T_4$ output:} Converged SCF energy and total wall-clock time (seconds) per molecule. KerneLDI SCF times range from $\sim$50~s (porphy) to $\sim$3000~s (crambin); speedup over GPU4PySCF should be $1.5\times$--$2.7\times$.
    \item \textbf{$T_5$ output:} Wall-clock time for each GPU count. Expected speedup at 64~GPUs: $\sim$6.3$\times$ relative to 8~GPUs (79\% parallel efficiency).
    \item \textbf{$T_6$ output:} MD trajectory file and per-step timing. Correctness: total energy should be conserved within SCF tolerance across steps.
\end{itemize}